\documentclass[a4paper,11pt]{article}
\pdfoutput=1
\usepackage{jcappub}
\usepackage{setspace}  

\usepackage{tikz,xcolor,hyperref}
\usepackage{amsmath, amssymb, amsthm, graphicx, epsfig, fancyhdr,epsfig, slashed}
\usepackage[normalem]{ulem}
\usepackage{tikzsymbols}
\usepackage{natbib}
\usepackage{float}
\usepackage{bm}
\usepackage{adjustbox}
\usepackage[utf8]{inputenc}
\usepackage[T1]{fontenc}

\usepackage[bitstream-charter]{mathdesign}
\usepackage[nointegrals]{wasysym} 

\DeclareSymbolFont{myletters}{OML}{ztmcm}{m}{it}
\DeclareMathSymbol{\uplambda}{\mathord}{myletters}{"15}
\DeclareMathSymbol{\upxi}{\mathord}{myletters}{"18}

\usepackage{amsfonts}

\renewcommand{\figurename}{FIGURE}

\usepackage{lmodern} 
\usepackage{amsmath}                                                    
\usepackage{amsthm}                                                     
\usepackage{amssymb}
\usepackage{multirow}
\usepackage[nameinlink]{cleveref}
\Crefname{figure}{Fig.}{Figs.}

\usepackage{adjustbox}
\usepackage{color, colortbl} 
\definecolor{Gray}{gray}{0.9}
\usepackage{ragged2e}
\usepackage{titlesec}
\usepackage[includemp,
paperwidth=21.5cm,
paperheight=29.7cm,
top=2.30cm,
bottom=2.3cm,
inner=3cm,
outer=2.0cm,
marginparwidth=0.3cm,
marginparsep=0.3cm]{geometry}
\newcolumntype{P}[1]{>{\centering\arraybackslash}p{#1}}
\newcolumntype{M}[1]{>{\centering\arraybackslash}m{#1}}
\usepackage{enumitem}
\usepackage{cellspace}
\usepackage{makecell}
\usepackage{setspace}
\usepackage[nameinlink]{cleveref}
\Crefname{figure}{Fig.}{Figs.}
\usepackage{subfig}
\usepackage{graphicx}
\usepackage{mathtools}
\usepackage{orcidlink}

\def\beq{\beq\begin{align}}
\def\eeq{\end{align}\eeq}

\def\beq{\begin{equation}\begin{align}}
\def\eeq{\end{align}\end{equation}}

\begin{document}
\begin{spacing}{1} 
\title{Primordial Black Holes and Scalar-induced Gravitational Waves in Radiative Hybrid Inflation }
\author[]{Adeela Afzal \orcidlink{0000-0003-4439-5342}$^{a\dagger}$,}
\author[]{Anish Ghoshal \orcidlink{0000-0001-7045-302X}$^{b*}$}

\affiliation[a]{ Department of Physics, \href{ https://ror.org/04s9hft57}{Quaid-i-Azam University, Islamabad, 45320, Pakistan}}
\affiliation[b]{ Institute of Theoretical Physics, Faculty of Physics,\\ University of Warsaw,
ul. Pasteura 5, 02-093 Warsaw, Poland}

\emailAdd{adeelaafzal555@gmail.com$^\dagger$}
\emailAdd{anish.ghoshal@fuw.edu.pl$^*$}

\abstract{We study the possibility that primordial black holes (PBHs) can be formed from large curvature perturbations generated during the waterfall phase transition due to the effects of one-loop radiative corrections of Yukawa couplings between the inflaton and a dark fermion in a non-supersymmetric hybrid inflationary model. We obtain a spectral index $n_s$, and a tensor-to-scalar ratio $r$, consistent with the current Planck data. Our findings show that the abundance of PBHs is correlated to the dark fermion mass $m_N$ and peak in the GW spectrum. We identify parameter space where PBHs can be the entire dark matter (DM) candidate of the universe or a fraction of it. Our predictions are consistent with any existing constraints of PBH from microlensing, BBN, CMB, etc. Moreover, the scenario is also testable via induced gravitational waves (GWs) from first-order scalar perturbations detectable in future observatories such as LISA and ET. For instance, with inflaton mass $m \sim 2\times 10^{12}$ GeV, $m_N \sim 5.4\times 10^{15}$ GeV, we obtain  PBHs of around $10^{-13}\, M_\odot$ mass that can explain the entire abundance of DM and predict GWs with amplitude $\Omega_{\rm GW}h^2$  $\sim 10^{-9}$ with peak frequency $f$ $\sim$ $0.1$ Hz in LISA. By explicitly estimating fine-tuning we show that the model has very mild tuning. We discuss successful reheating at the end of the inflationary phase via the conversion of the waterfall field into standard model (SM) particles. We also briefly speculate a scenario where the dark fermion can be a possible heavy right-handed neutrino (RHN) which is responsible for generating the SM neutrino masses via the seesaw mechanism. The RHN can be produced due to waterfall field decay and its subsequent decay may also explain the observed baryon asymmetry in the universe via leptogenesis. We find the reheat temperature $T_R\lesssim5\times10^9$~GeV that explains the matter-anti-matter asymmetry of the universe.} 
 
\maketitle

\section{Introduction}
\label{sec:intro}


The seeds of the structure of the universe can be generated by quantum fluctuations of inflaton or curvaton during inflation.
The amplitude of curvature perturbations is of the order of $10^{-5}$ at the cosmic microwave background (CMB) scale~\cite{Planck:2018vyg}, whereas larger curvature perturbations may be generated at a smaller scale~\cite{Ivanov:1994pa, GarciaBellido:1996qt, Kawasaki:1997ju, Yokoyama:1998pt, Garcia-Bellido:2017mdw, Hertzberg:2017dkh}. 
Observations of the supermassive ~\cite{LyndenBell:1969yx, Kormendy:1995er} and stellar-mass black hole (BH) merger events by gravitational wave (GW) detectors~\cite{LIGOScientific:2016dsl, LIGOScientific:2021djp} imply the existence of primordial black holes (PBHs). These PBHs are generated via the collapse of high-dense regions~\cite{Hawking:1971ei,Carr:1974nx,Carr:1975qj}.%
\footnote{
There are some other scenarios to have a PBH formed: cosmic strings~\cite{Hawking:1987bn, Garriga:1992nm, Caldwell:1995fu}, bubble collisions~\cite{Hawking:1982ga}, domain walls~\cite{Garriga:1992nm, Khlopov:2008qy, Garriga:2015fdk, Deng:2016vzb,Gouttenoire:2023ftk,Gouttenoire:2023gbn}, and collapse of vacuum bubbles~\cite{Garriga:2015fdk, Deng:2017uwc, Deng:2018cxb, Deng:2020mds, Conaci:2024tlc} or from multiple fields \cite{Pi:2017gih, Chen:2023lou}. }
For other mechanism of PBH and GW production see refs. ~\cite{ Teimoori:2021thk, Ashrafzadeh:2023ndt, Heydari:2023xts}.
The PBH is also a candidate for dark matter (DM)~\cite{Chapline:1975ojl} if its mass is within $10^{17\text{--}23} $g (see, e.g., refs.~\cite{Carr:2016drx,Inomata:2017okj,Inomata:2017vxo}). 
Such large curvature perturbations can be generated if the inflaton or a spectator field goes through a very flat potential or has double potential like shapes \cite{Ghoshal:2024hfk} during inflation (see ref.~\cite{Escriva:2022duf} for a recent review on PBHs). 

However, going beyond the single-field inflationary scenarios which involve some degrees of fine-tuning \cite{Cole:2023wyx}, one may also generate large curvature perturbations in the hybrid inflation models. Here, inflation ends with a waterfall phase transition~\cite{Linde:1993cn} with the natural possibility of being embedded in grand unification schemes. Particularly, it is well known that the Grand Unified Theory (GUT) scale $M_{\rm GUT}$ for such waterfall phase transition corresponds to the observed amplitude of primordial scalar fluctuations as seen in the CMB~\cite{Dvali:1994ms}. 

Well studied in the literature, the waterfall field potential may be flat leading to the generation of large curvature perturbations at small scales, since the waterfall transition happens at the later stages of inflation~\cite{Clesse:2010iz,Kodama:2011vs,Mulryne:2011ni,Afzal:2024hwj}. 
These large density perturbations which subsequently lead to large curvature perturbations may collapse into a PBH and give us observable effects. Particularly following refs.~\cite{Clesse:2015wea,Kawasaki:2015ppx}, one may understand that more often than not it led to the overproduction of PBHs of astrophysical size\footnote{See also refs.~\cite{Choi:2021yxz,Ijaz:2023cvc,Ijaz:2024zma} for PBH from hybrid inflation.}. However, with a slightly modified waterfall field potential, this can be avoided and PBH abundance can be controlled with detectable scalar-induced GW signals in the next generation GW observatories \cite{Braglia:2022phb,Roshan:2024qnv,Yuan:2021qgz}. The models discussed in ~\cite{Clesse:2015wea,Kawasaki:2015ppx,Stamou:2021qdk} also suffer with the problem of initial conditions that need to be set to satisfy current constraints. This nonetheless can be avoided by considering $\alpha$-attractor models \cite{Kallosh:2022ggf}. Since there are several large uncertainties present in the
critical process of PBH formation estimates, we study it in less detail. On the other hand, we study in detail the characterization of the stochastic gravitational wave background (SGWB) induced at the second-order by the large curvature perturbations (at small length scales) during horizon re-entry in
the radiation dominated era. These GWs will be used to test the model involving dark fermions and seesaw physics involving radiative corrections to the hybrid inflation potential. Careful attention has been paid to the consistency with large-scale measurements of the CMB radiation anisotropy from the latest
Planck/BICEP/Keck Array release \cite{BICEPKeck:2022mhb}. This involves producing
a peak at the small scale, but keeping the amplitude of the power spectrum small at the CMB scale. A best Planck fit spectral index is obtained without the problem of fine-tuning, which is usually the case for single-field inflationary scenarios. We show from our fine-tuning estimates involving the parameter space of dark fermion masses and quartic couplings that the fine-tuning in the model is mild and largely reduced from that of the single-field scenarios by explicitly estimating the fine-tuning quotient. 

We investigate a non-supersymmetric particle physics framework where the inflaton field is responsible for dark fermion mass generation \cite{Rehman:2009wv}. Because of the Yukawa coupling between the two, the inflaton field receives radiative quantum corrections which plays a vital role in setting the inflaton dynamics such that the power spectrum gets enhanced at smaller scales and sets the amplitude of the curvature spectrum at the CMB scale. However, in the hybrid inflationary scenario, there is further growth of perturbation during the waterfall transition which leads to the formation of PBH that can be a dark matter candidate and also induces tensor perturbations at second order propagating as SGWB.  We show the second-order tensor perturbations propagating as GWs with amplitude $\Omega_{\rm GW}h^2$  $\sim 10^{-9}$ and peak frequency f $\sim$ $0.1$ Hz for LISA and $\Omega_{\rm GW}h^2 \sim 10^{-11}$ and peak frequency of $\sim$ $10$ Hz in ET. Production of PBH of mass around $ 10^{-13} M_\odot$ explains the entire abundance of DM in the universe that corresponds to the inflaton mass $m \sim 2\times 10^{12}$ GeV and dark fermion mass $m_N \sim 5.4\times 10^{15}$ GeV.
 This novel DM candidate is also a signature of the scale of dark fermion physics involving inflationary cosmology. We furthermore speculate that
 the dark fermions can be the possible candidate for a right-handed neutrino (RHN) responsible for generating SM neutrino mass via the seesaw mechanism and leaving imprints in the scalar power spectrum and GWs. This RHN may also explain the baryon asymmetry of the universe through baryogenesis via leptogenesis.

\textit{This paper is structured as follows:} In \cref{alphatt}, we present $\alpha$-attractor radiative hybrid inflation with dark fermions. In \cref{scalarpert}, we present the power spectrum of scalar perturbations and formation mechanism for PBHs. In \cref{SIGWs}, we present the power spectrum for scalar-induced GWs and calculate the signal-to-noise ratio for future planned detectors and correlation between the model parameters. In \cref{finetuneapp}, we give the fine-tuning estimate and compare it with a single field and standard hybrid inflation and discuss the reheating. In \cref{RHN}, we explain that our proposed dark fermion can also be a RHN, we discuss the seesaw mechanism, reheating and non-thermal leptogenesis and conclude in \cref{concl}. The paper also contains \cref{appGW} where we briefly discuss our model in the Pulsar Timing Arrays (PTAs) band.

\section{$\alpha$-attractor Radiative Hybrid Inflation with Dark Fermions}
\label{alphatt}
We explore the radiative hybrid inflation in the context of an $\alpha$-attractor model where the kinetic part of the inflaton field $\phi$ is modified\footnote{For simplicity waterfall field kinetic term is not modified in the present set-up following ref.\cite{Kallosh:2022ggf} but can be introduced for detailed investigations.}. Working with Einstein-Hilbert gravity action, the relevant terms for the inflaton field $\phi$, the waterfall field $\psi$ and a dark fermion $N$ in the Lagrangian density can be written as,
\begin{align}
\label{Lagrang}
    \mathscr{L}\simeq &\dfrac{\left(\partial^\mu \phi\right)^2}{2\left(1-\dfrac{\phi^2}{6\,\alpha}\right)^2}+\dfrac{\left(\partial^\mu\psi\right)^2}{2}+\dfrac{i}{2}\Bar{N}\gamma^\mu\partial_\mu N-\kappa^2\left(M^2-\dfrac{\psi^2}{4}\right)^2-\dfrac{1}{2}\,m^2\,\phi^2\\\notag
   & -\dfrac{\lambda^2}{4}\,\phi^2\,\psi^2 -\dfrac{1}{2}y\,\phi\,\Bar{N}\,N-\dfrac{1}{2}Y\,\psi\,\Bar{N}\,N-\dfrac{1}{2}m_N\,\Bar{N}\,N.
\end{align}
The Yukawa interaction between $\phi$ and $N$ results in one-loop radiative corrections. The hybrid inflationary potential including one-loop radiative corrections and a linear term arising from the interaction between $\psi$ and $N$ can be written as:
\begin{align} \label{hpoten}
    V(\psi,\phi)&= \kappa^2\left(M^2-\dfrac{\psi^2}{4}\right)^2+\dfrac{1}{2}\,m^2\,\phi^2+\dfrac{\lambda^2}{4}\,\phi^2\,\psi^2+b^3\, \psi  + V_\text{loop}(\phi),
\end{align}
where the one-loop radiative correction is given according to Ref. \cite{Rehman:2009wv},
\begin{align}
\label{Vloop}
    V_\text{loop}(\phi)&=\dfrac{1}{64\,\pi^2}\left[m^4\,\text{ln}\left(\dfrac{m^2}{\mu^2}\right)+\dfrac{\lambda^4}{4}\left(\phi^2-\phi_c^2\right)^2\,\text{ln}\left(\dfrac{\lambda^2\lvert\phi-\phi_c\rvert^2}{2\,\mu^2}\right)\right. \\\notag 
&\left.-2\,\left(m_N+y\,\phi\right)^4\,\text{ln}\left(\dfrac{m_N+y\,\phi}{\mu}\right)^2\right].
\end{align}
Here, $M$, $m$ (inflaton mass) and $b$, are the dimensional mass parameters, while $\kappa$, $\lambda$ and $y$ are dimensionless couplings. The parameters $\mu$ and $m_N$ are the cutoff scale and the dark fermion mass, respectively. 
The linear term in $\psi$ in the potential \cref{hpoten} can arise in many instances, e.g., from the non-perturbative generation of condensates with respect to some dark sector interactions \cite{Barenboim:2008ds,Bardeen:1989ds,Bhatt:2008hr,Bhatt:2009wb,Barenboim:2010nm}, or due to gravitational interactions~\cite{Barenboim:2010db,Dvali:2016uhn}, or from specific SUSY related motivations~\cite{Rehman:2009nq,Dvali:1994ms,Copeland:1994vg,Lazarides:2001zd,Lyth:1998xn,Takahashi:2010ky}~and from several other scenarios involving instanton physics \cite{Davoudiasl:2020opf,Iso:2014gka,Kaneta:2017lnj}. Here, we do not go into such detailed scenarios but instead, use it as a phenomenological term since our analysis remains applicable to all such scenarios. Moreover, the linear term does not affect the PBH and GW cosmology, except to not overproduce the PBH. We will see later that it can be tuned accordingly to choose the relevant benchmark points (BPs) that we study.

\begin{figure}[t]
    \centering
    \includegraphics[width=0.7\linewidth]{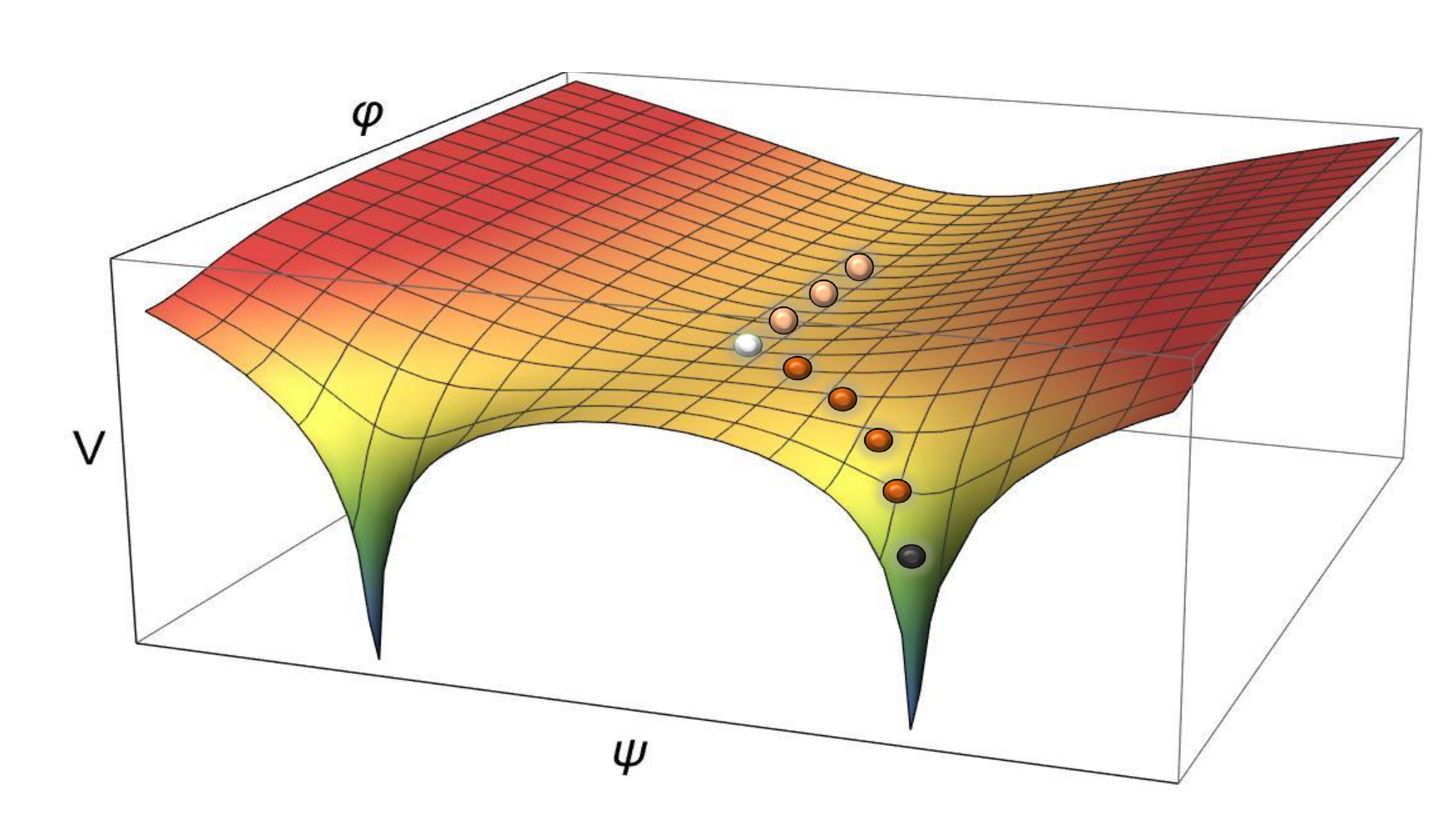}
    \caption{ \it A schematic picture of radiative hybrid inflation potential. The light orange bullet points show the inflationary trajectory of the inflaton $\varphi$, the white bullet is the critical point and the dark orange bullets show the waterfall regime. Inflation continues for some e-folds after the critical point during which very large curvature perturbations are generated. }
    \label{fig:potential}
\end{figure}

Along the valley, the waterfall field stabilizes at $\psi=0$ as long as $\phi$ is larger than the critical field value $\phi_c$,  see \cref{fig:potential}. After $\phi$ crosses $\phi_c$ the field $\psi$ falls into one of
the two minima of the potential at $\psi \simeq \pm M$ depending on the sign of the coefficient of the linear term $b$. The radiative hybrid potential \cref{hpoten}, 
in terms of the canonically normalized inflaton field \cite{Kallosh:2022ggf}, $\phi\rightarrow \sqrt{6\,\alpha} \text{Tanh}\left(\varphi/\sqrt{6\,\alpha}\right)$, can now be written as,
\begin{align}
\label{canonpoten}
V(\psi,\varphi) &= \kappa^2\left(M^2-\dfrac{\psi^2}{4}\right)^2+\dfrac{1}{2}\,\left(m^2+\dfrac{\lambda^2}{2}\psi^2\right)\,\left(\sqrt{6\,\alpha}\, \text{Tanh}\left(\dfrac{\varphi}{\sqrt{6\,\alpha}}\right)\right)^2 +b^3\, \psi \\\notag
&+V_\text{loop}\left(\sqrt{6\,\alpha}\, \text{Tanh}\left(\varphi/\sqrt{6\,\alpha}\right)\right).
\end{align}
 The schematic view of hybrid potential is shown in \cref{fig:potential}. The mass squared of the waterfall field at $\psi=0$ is,
\begin{align}
\label{waterfallmass}
M_\psi^2=\left(-\kappa^2\,M^2+\dfrac{1}{2}\left(\lambda\,\sqrt{6\,\alpha}\,\text{Tanh}\left(\dfrac{\varphi}{\sqrt{6\,\alpha}}\right)\right)^2\right).
\end{align}
In this paper, we assume that $(\lambda\,\sqrt{6\,\alpha})^2/2 > \kappa^2\,M^2$ such that $M_\psi^2>0$ at large $\varphi>\varphi_c$ to stabilize the inflationary trajectory at $\psi = 0$, where,
\begin{align}
\label{phicrit}
    \text{Tanh}^2\left(\dfrac{\varphi_c}{\sqrt{6\,\alpha}}\right)=\dfrac{\kappa^2\,M^2}{3\,\alpha\,\lambda^2}.
\end{align}
During inflation, as long as $\varphi\lesssim\varphi_c$, the effective mass square of $\psi$ becomes negative which gives rise to tachyonic instability that will grow the curvature perturbations. These growing perturbations will enhance the scalar power spectrum at small scales and upon horizon re-entry the collapse of large density fluctuations produces the PBHs. Due to the linear term in the potential \cref{canonpoten} the field $\psi$ will not relax exactly at $\psi=0$ but will be displaced depending upon the sign. of the coefficient of the linear term $b$. In this way, we can inflate the unnecessary topological defects and control the peak of the power spectrum at small scales to avoid PBH overproduction.  

The slow-roll parameters are given by, \cite{Chatterjee:2017hru},
\begin{align}
    \epsilon_V &= \dfrac{m_\text{Pl}^2}{2}\left(\dfrac{\partial_\xi V}{V}\right)^2,\,\,\,\,\,\eta_V = m_\text{Pl}^2\left(\dfrac{\partial_\xi^2 V}{V}\right), \\\notag
    \delta_V^2 & = m_\text{Pl}^4 \left(\dfrac{\partial_\xi V\,\partial_\xi^3 V}{V^2}\right),\,\,\,\,\, \sigma_V^3= m_\text{Pl}^6 \left(\dfrac{(\partial_\xi V)^2\,\partial_\xi^4 V}{V^3}\right).
\end{align}
Here, $m_\text{Pl}\simeq 2.43 \times 10^{18}$ 
 GeV is the reduced Planck mass and the subscript $\xi=\{\varphi,\psi\}$ indicates the field derivative. In the slow roll limit, along the valley at the pivot scale, the spectral index $n_s$, its running and running of the running, are given by \cite{Chatterjee:2017hru},
\begin{align}
    n_s &= 1-6\,\epsilon_V + 2\,\eta_V,\,\,\,\,\, \text{d}n_s/\text{d}\ln k=16\,\epsilon_V\,\eta_V-24\,\epsilon_V^2-2\,\delta_V^2,\\\notag
    \text{d}^2n_s/\text{d}\ln k^2 &=-192\,\epsilon_V^3+192\,\epsilon_V^2\,\eta_V-32\,\epsilon_V\,\eta_V^2-24\,\epsilon_V\delta_V^2+2\,\eta_V\delta_V^2+2\sigma_V^3.
\end{align}
\begin{figure}[t]
    \centering
    \includegraphics[width=0.7\linewidth]{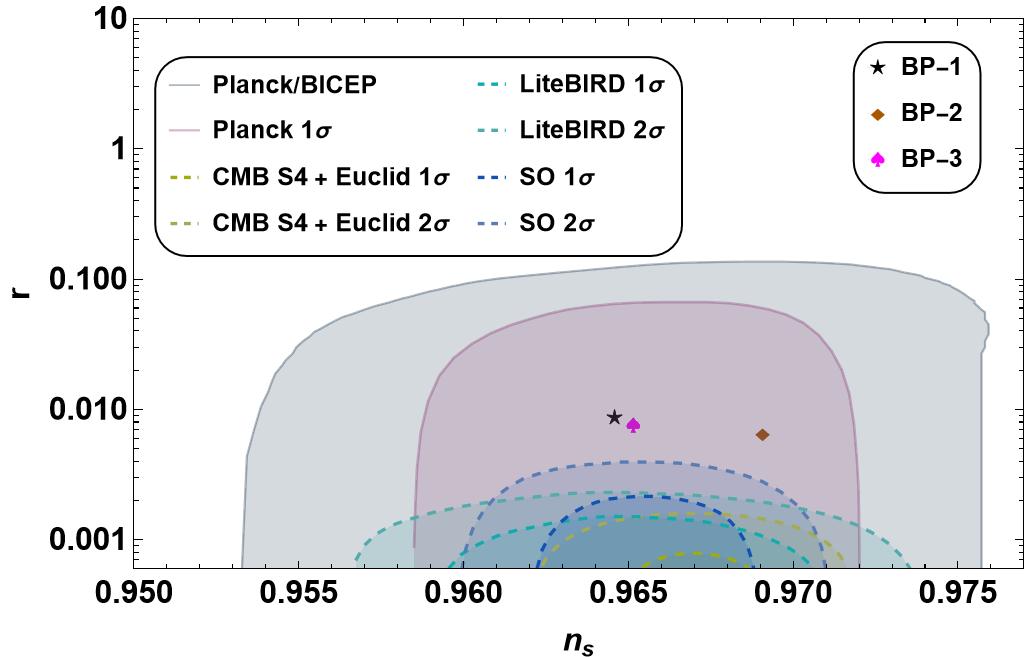}
    \caption{ \it Tensor-to-scalar ratio $r$ vs. scalar spectral index $n_s$ for the corresponding parameter sets given in \cref{parmsets}. The solid contours are the current Planck bounds \cite{Planck:2018jri}, Planck/BICEP \cite{BICEP:2021xfz,Planck:2018vyg,BICEPKeck:2022mhb} and the dashed shaded region indicates the future proposed experiments (LiteBIRD, CMB S4-Euclid, Simons Observatory (SO)) \cite{laureijs2011euclid, SimonsObservatory:2018koc, LiteBIRD:2020khw}. }
    \label{fig:rns}
\end{figure}
The central measurements by Planck 2018 \cite{Planck:2018vyg} in the $\Lambda$CDM model are; $n_s = 0.9647 \pm 0.012$, $\text{d}n_s/\text{d}\ln k =  0.0011 \pm 0.0099$ and $ \text{d}^2n_s/\text{d}\ln k^2 = 0.009 \pm 0.012$. The tensor to scalar ratio $r=16\,\epsilon_V < 0.036$ at $95\%$ C.L. All these values are measured at the pivot scale, $k_\star=0.05\,\text{Mpc}^{-1}$. For BP-$1$ in \cref{parmsets}, we find $\text{d}n_s/\text{d}\ln k \simeq -0.0001373$ and $ \text{d}^2n_s/\text{d}\ln k^2\simeq 0.00002961$.
The prediction of $n_s$ and $r$ given in \cref{parmsets} is shown in \cref{fig:rns} within $1\sigma$ bound of recent Planck results.
\begin{table}[H]
	  \fontsize{14pt}{15pt}
	\caption{ \it Benchmark points (BPs) for model parameters } 
	\centering 
	\begin{adjustbox}{max width=\columnwidth}
		\begin{tabular}{|M{2.2cm} |M{1.4cm} |M{2.2cm} |M{2.2cm} |M{2.6cm}|M{2.7cm}|M{1.7cm}|M{1.5cm}|M{1.5cm}|M{1cm}|M{1cm}|M{1.5cm}|}
			\hline
			\bf{Model}& \bf{$M/m_\text{Pl}$}&\bf{$m/m_\text{Pl}$}&\bf{$\lambda$}&\bf{$\kappa$}&\bf{$b$}&\bf{$\sqrt{\alpha}/m_\text{Pl}$}&\bf{$\varphi_i/m_\text{Pl}$}&\bf{$\psi_i/m_\text{Pl}$}
			\\ [0.6ex] 
			\hline\hline
			$\text{BP-1}$ &$1.35$  &$8.40\times 10^{-7}$  & $2.83\times 10^{-6}$ & $ 2.181\times 10^{-6} $ &$-3.15\times 10^{-6}$ &$1$ & $3.5$ & $0$\\
			\hline 
   $\text{BP-2}$ &$1.46$  &$6.10\times 10^{-7}$  & $7.10\times 10^{-6}$ & $ 1.450\times 10^{-6} $ &$-3.50\times 10^{-6}$ &$1$ & $3.5$ & $0$\\
			\hline
   $\text{BP-3}$ &$1.34$  &$7.70\times 10^{-7}$  & $2.83\times 10^{-6}$ & $ 1.800\times 10^{-6} $ &$-3.15\times 10^{-6}$ &$1$ & $3.5$ & $0$\\
			\hline
		\end{tabular}
	\end{adjustbox}
	\label{parmsets}
\end{table}
\begin{table}[H]
	 \fontsize{14pt}{15pt}
	\centering 
	\begin{adjustbox}{max width=\columnwidth}
		\begin{tabular}{|M{1.8cm}|M{2.3cm} |M{2.3cm} |M{2.3cm} |M{1.4cm}|M{1cm}|M{1.3cm}|M{1.5cm}|M{1.5cm}|M{1.5cm}|}
			\hline
			\bf{Model}& \bf{$\mu/m_\text{Pl}$}&\bf{$m_N/m_\text{Pl}$}&\bf{$y$}&\bf{$\phi_c/m_\text{Pl}$}&\bf{$N_k$}&\bf{$n_s$}&\bf{$r$}&\bf{$\varphi_\text{e}/m_\text{Pl}$}&\bf{$\psi_\text{e}/m_\text{Pl}$}
			\\ [0.6ex] 
			\hline\hline
			$\text{BP-1}$ &$2.94 \times 10^{-6}$  &$2.23 \times 10^{-3}$  & $2.12\times 10^{-6}$ & $ 1.70 $ & $57$ & $0.9645$ & $0.00873$ & $0.27$ & $1.37$ \\
			\hline 
   	$\text{BP-2}$ &$2.13 \times 10^{-6}$  &$1.15 \times 10^{-3}$  & $5.02\times 10^{-6}$ & $ 0.42 $ & $55$ & $0.9690$ & $0.00657$ & $0.0018$  & $2.05$ \\
			\hline
   	$\text{BP-3}$ &$3.40 \times 10^{-7}$  &$2.23 \times 10^{-6}$  & $5.10\times 10^{-4}$ & $ 1.32 $ & $57$ & $0.9651$ & $0.00756$& $0.079$ & $1.62$\\
			\hline
		\end{tabular}
	\end{adjustbox}
\end{table}

\medskip


To avoid eternal inflation, $M$ should be $O(1)$ \cite{Braglia:2022phb}. The parameter $m$ along with the loop corrections controls the amplitude of the plateau in the valley. We choose $y\gtrsim \lambda/\sqrt{2}$ and set $m_N$ as a free parameter such that the dominant contribution to the potential comes from the last term in \cref{Vloop} thus making it radiatively generated. We set the normalization scale $\mu$ so the log term remains positive. The coupling $\lambda$ determines the number of e-folds in the waterfall regime and $\kappa$ will fix the amplitude of the power spectrum at the pivot scale, that is, $A_s(k_\star)\simeq 2.24\times10^{-9}$.
 Taking into account all these constraints we define the BPs in \cref{parmsets} for the potential in \cref{canonpoten} and study the production of PBH and the scalar-induced SGWB in the subsequent sections. The subscripts $i$ and $e$ in \cref{parmsets} refer to the values at the start and end of inflation. The presence of the linear term makes the potential unbounded close to the critical point since the coefficient $b$ is very small and for larger field values it does not play a significant role.
\section{Scalar Perturbations and Primordial Black Hole Formation}
\label{scalarpert}

Let us explain the generation and evolution of scalar perturbations during inflation.

\subsection{Scalar Spectra}

\begin{figure}[t]
    \centering
    \includegraphics[width=0.7\linewidth]{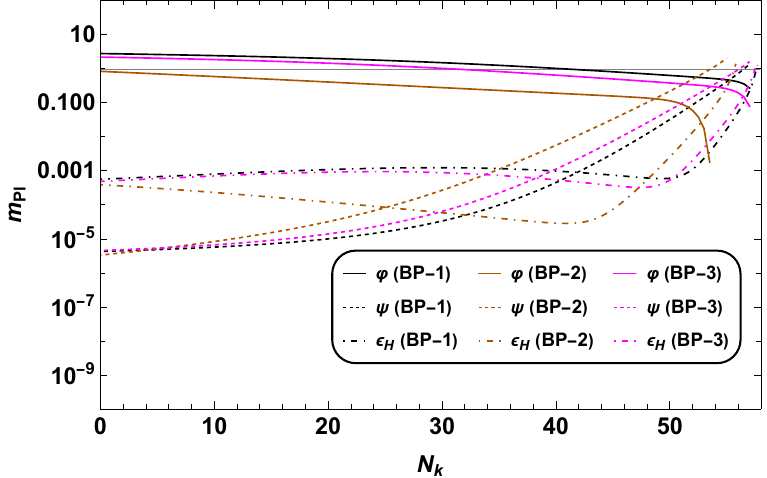}
    \caption{ \it Field and the slow-roll parameter $\epsilon_H\equiv -H^{'}/H$ evolution with the number of e-folds from pivot scale to the end of inflation. We evaluate solving the full background \cref{bac} using potential in \cref{canonpoten} for the BPs in \cref{parmsets}.}
    \label{fig:phipsi}
\end{figure}

The background equations of motion in the number of e-fold times are given by \cite{Clesse:2013jra},
\begin{align}
\label{bac}
	\varphi^{''}+\left(\dfrac{H^{'}}{H}+3\right)\varphi^{'}+\dfrac{V_\varphi}{H^2}=0,\,\,\,\,\,\,\,\,\,\,\,\,\,\,
	\psi^{''}+\left(\dfrac{H^{'}}{H}+3\right)\psi^{'}+\dfrac{V_{\psi}}{H^2}=0,
\end{align}
where $H$, the Hubble rate is defined to be, $H^2=2\,V/(6-\varphi^{'2}-\psi^{'2})$.
Here, prime is the derivative with respect to the number of e-folds and $V_\xi=dV/d\xi$ where $\xi=\{\varphi,\psi\}$. The evolution of the fields is shown in \cref{fig:phipsi}, where the background is calculated until the end of inflation, that is, when $\epsilon_H\equiv-H^{'}/H=1$. For different BPs given in \cref{parmsets}, the critical point shifts and affects the number of e-folds in the waterfall regime. 
Usually, the evolution of quantum fluctuations is affected due to the back-reaction of long modes. Although, for detailed analysis one needs to perform the lattice simulations which is beyond the scope of the present work. As a naive criterium, which should anyway give a good estimate of when back-reaction is negligible, we may see that the background energy density; $H^2\,\phi^{'2}/2+m^2\,\phi^2/2\ll M^4$, for the BPs we chose in \cref{parmsets}. This enables us to disregard the backreaction on metric as discussed in detail in  \cite{GarciaBellido:1996qt, Braglia:2022phb}

The scalar perturbations of the Friedmann–Lemaître–Robertson–Walker (FLRW) 
metric in longitudinal gauge can be expressed as \cite{Clesse:2013jra};
\begin{align}
	ds^2=a(\tau)^2\left[(1+2\,\Phi_{\text{B}})d\tau^2+\left[(1-2\,\Psi_{\text{B}})\delta_{ij}+\dfrac{h_{ij}}{2}\right]dx^idx^j\right],
\end{align}
where $\Phi_{\text{B}}$ and $\Psi_{\text{B}}$ are the Bardeen potentials, $h_{ij}$ is the transverse-traceless tensor metric perturbation i.e. $h_i^i= 0 = h^j
_{i,j}$. The conformal time $\tau$, is related to cosmic time, $dt=a\,d\tau$, $a$ being the scale factor. Working in the conformal Newtonian gauge we set, $\Phi_{\text{B}}=\Psi_{\text{B}}$. Following the dynamics given in \cite{Ringeval:2007am, Clesse:2013jra} to calculate the power spectrum,
the Klein-Gordon equation to evaluate scalar perturbations is given by,
\begin{align}
	\delta \xi_i^{''}+(3-\epsilon)\delta \xi_i^{'}+\sum_{j=1}^{2}\dfrac{1}{H^2}V_{\xi_i \xi_j}\delta \xi_j+\dfrac{k^2}{a^2H^2}\delta \xi_i=4\Phi^{'}_{\text{B}}\,\xi^{'}_i-\dfrac{2\,\Phi_{\text{B}}}{H^2}V_{\xi_i}.
\end{align}
Here $\xi$ with subscript $(i, j)$ refers to the fields ($\varphi$, $\psi$), $k$ is the comoving wave vector, the equation of motion for $\Phi_{\text{B}}$ is given by,
\begin{align}
	\Phi^{''}_{\text{B}}+(7-\epsilon)\,\Phi^{'}_{\text{B}}+\left(\dfrac{2\,V}{H^2}+\dfrac{k^2}{a^2H^2}\right)\Phi_{\text{B}}+\dfrac{V_{\xi_i}}{H^2}\,\delta \xi_i=0.
\end{align} 

The initial conditions (i.c) for field perturbations in e-fold time are given as,
\begin{align}
	\delta \xi_{i,\text{i.c}}=\dfrac{1}{a_{\text{i.c}}\sqrt{2 k}},\,\,\,\,\,\,\,\,\,\,\,\,\,\,
	\delta \xi_{i,\text{i.c}}^{'}=-\dfrac{1}{a_{\text{i.c}}\sqrt{2 k}} \left(1+\iota\dfrac{k}{a_{\text{i.c}}H_{\text{i.c}}}\right).
\end{align}
The initial conditions for the Bardeen potential and its derivative are given by,
\begin{align}
	\Phi_{\text{B,i.c}}=\sum_{i=1}^{2}\dfrac{\left(H_{\text{i.c}}^2 \xi_{i,\text{i.c}}^{'}\delta \xi_{i,\text{i.c}}^{'}+\left(3H_{\text{i.c}}^2 \xi_{i,i.c}^{'}+V_{\xi_i,\text{i.c}}\right)\delta \xi_{i,\text{i.c}}\right)}{2H_{\text{i.c}}^2\left(\epsilon_{\text{i.c}}-\dfrac{k^2}{a_{\text{i.c}}^2H^2_{\text{i.c}}}\right)},\,\,\,\,
	\Phi^{'}_{\text{B,i.c}}=\sum_{i=1}^{2}\dfrac{\xi_{i,\text{i.c}}^{'}\delta \xi_{i,\text{i.c}}}{2}-\Phi_{\text{B,i.c}}.
\end{align}
The scalar power spectrum $P_s(k)$ is defined as,
\begin{align}
\label{Powspec}
	P_s(k)=\dfrac{k^3}{2\pi^2}\left|\zeta\right|^2=\dfrac{k^3}{2\pi^2}\left|\Phi_{\rm{B}}+\dfrac{\sum_{i=1}^{2}\xi^{'}_i\delta \xi_i}{\sum_{j=1}^{2}\xi^{'2}_j}\right|^2.
\end{align}
 \begin{figure}[t]
    \centering
    \includegraphics[width=0.471\linewidth,height=5.7cm]{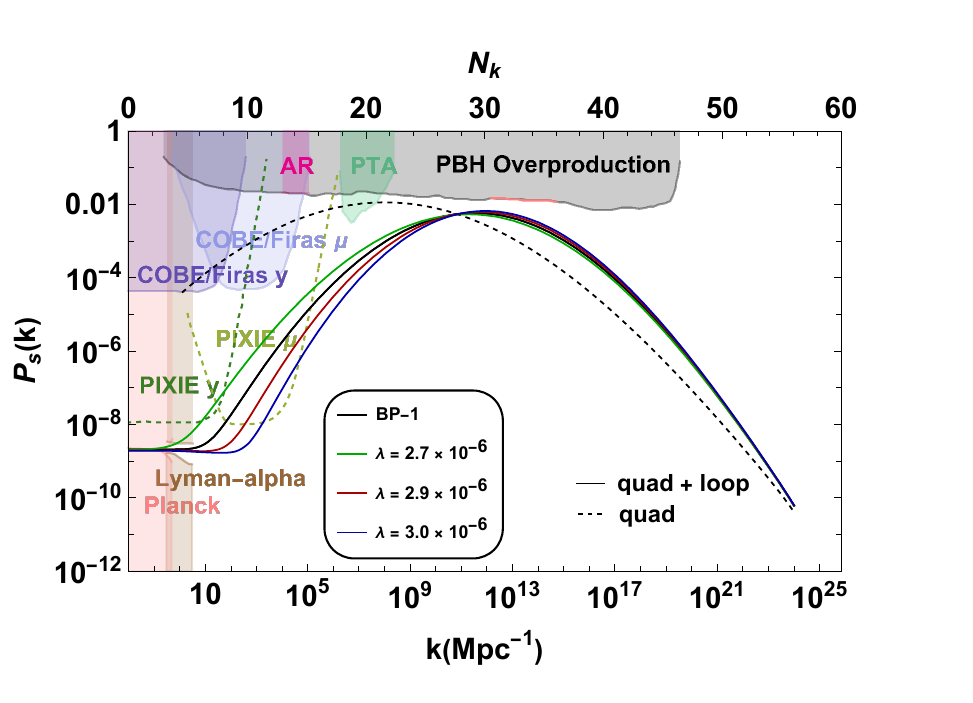}
    \quad
    \includegraphics[width=0.471\linewidth,height=5.7cm]{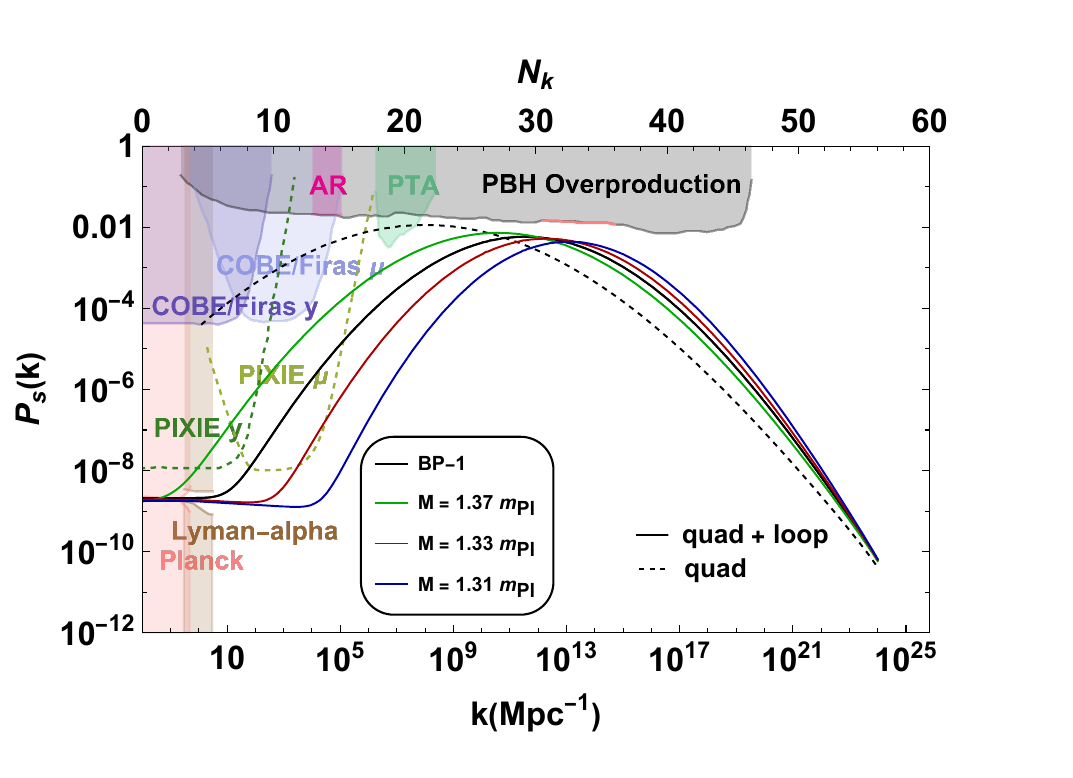}
      \quad
    \includegraphics[width=0.471\linewidth,height=5.7cm]{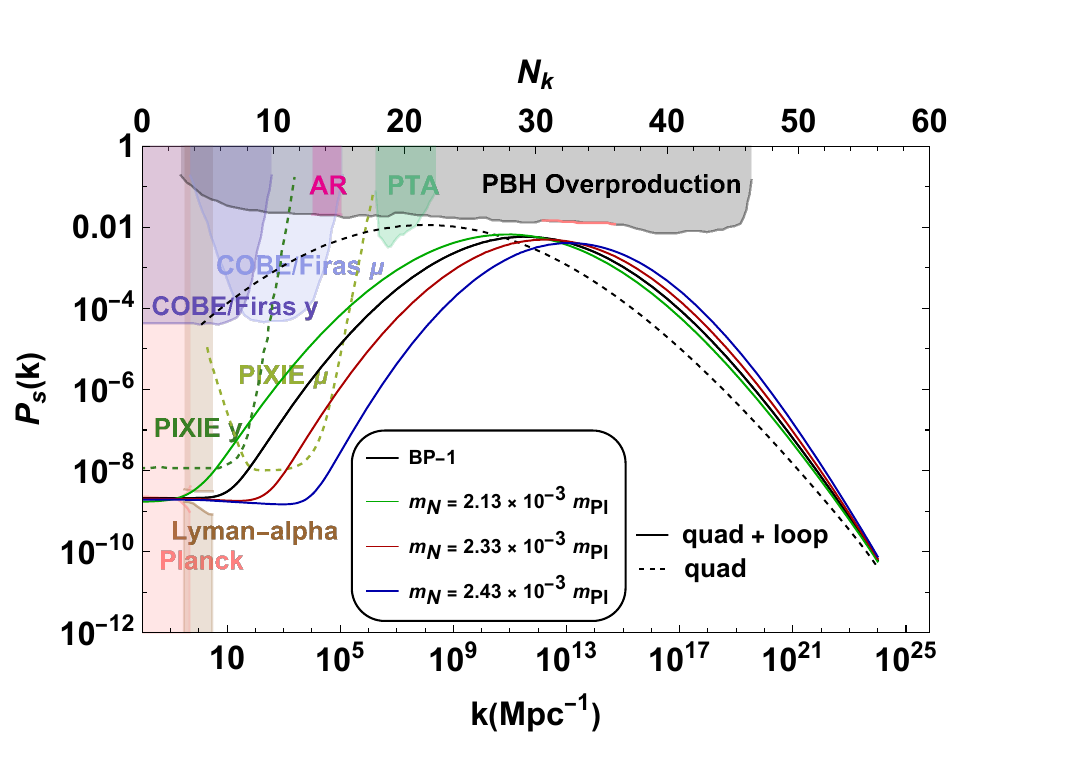}
      \quad
    \includegraphics[width=0.471\linewidth,height=5.7cm]{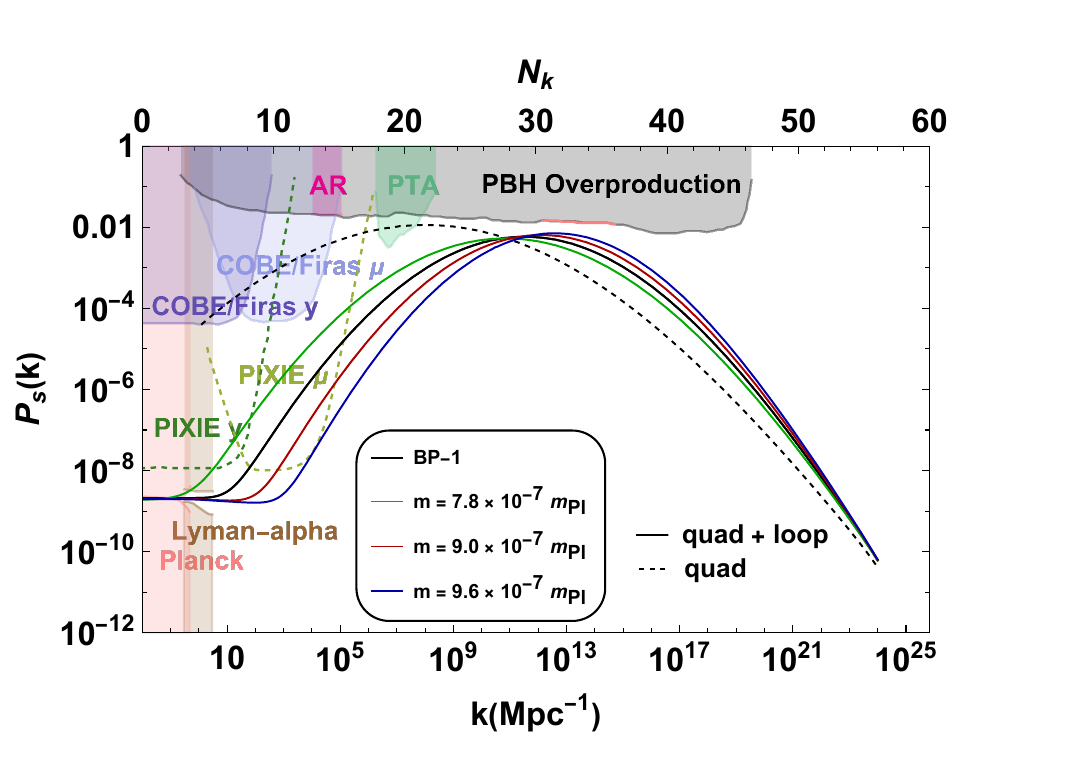}
          \quad
    \includegraphics[width=0.471\linewidth,height=5.7cm]{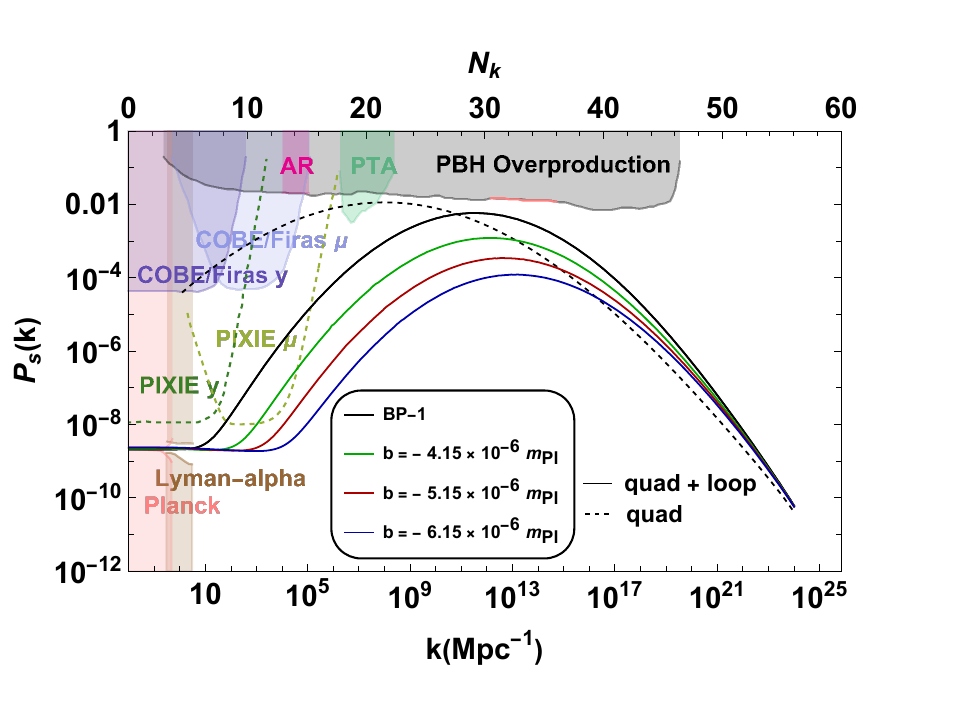}
              \quad
    \includegraphics[width=0.471\linewidth,height=5.7cm]{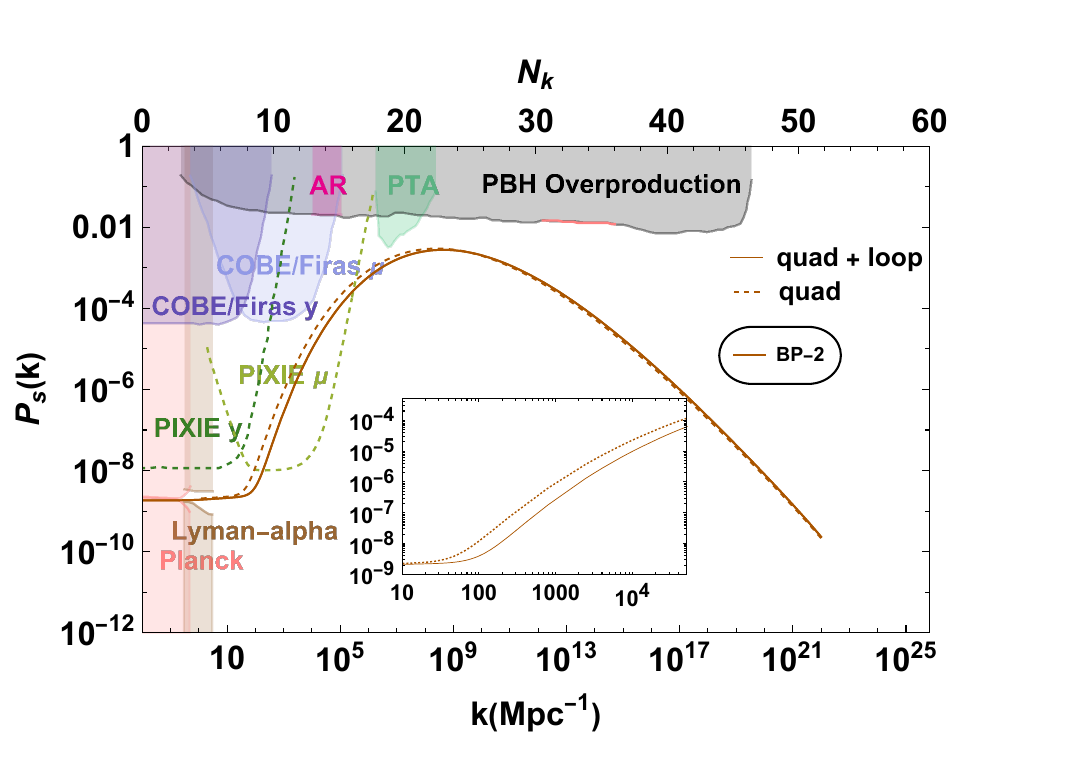}
    \caption{ \it  Power spectrum from pivot scale to the end of inflation by solving the exact scalar perturbation equations for BP given in \cref{parmsets}. The shaded region corresponds to the constraints from the present (solid) and future (dashed) experiments, see the main text for the details. The orange line in PBH overproduction bound represents the window where PBHs can explain the DM in totality.}
    \label{fig:PS_k}
\end{figure}

In this paper, we present an exact power spectrum (numerically solved) for the potential \cref{canonpoten} and the
the resulting spectrum is shown in  \cref{fig:PS_k} from the pivot scale to the end of inflation. One can see from \cref{fig:PS_k} that for a certain choice of the inflaton mass which is large (see BP-$1$ in \cref{parmsets}), it ruins the power spectrum predictions and runs into severe constraints at the CMB measurement. However, for this same choice, if the inflaton is coupled to the dark fermion $N$, then the quantum radiative loop corrections rectify the power spectrum and give a Planck consistent $n_s\simeq0.964$ at the CMB scale. For BP-$2$, without loop corrections, $n_s\simeq 0.974$ but with loop corrections, we obtain $n_s=0.969$ which is consistent with the measurements, see the inset of the bottom right in \cref{fig:PS_k}. 
Note that the duration of the waterfall is controlled by vev $M$. To avoid an abrupt end or a longer period of inflation, $M$ should be of the order $1$ in Planckian units. As the value of $M$ increases, we expect the duration of the waterfall transition to increase, which enhances the curvature perturbation $P_{s}(k)$. As in \cref{fig:PS_k}, the top right panel clearly depicts that the larger $M$ enhances the waterfall phase and the peak of the spectrum is shifted towards the left and the smaller value reduces the waterfall phase and the peak of the spectrum is moved towards the right.

We illustrate for comparison the current bounds from Planck \cite{Planck:2018jri}, Lyman-alpha \cite{Bird_2011}, PIXIE \cite{A_Kogut_2011}, COBE/Firas \cite{Fixsen:1996nj}, PTA \cite{Byrnes:2018txb}, the PBH overproduction that is related to the critical density $\delta_c$ which we discuss later and a window where PBH as the entire DM candidate of the universe (see orange line in the PBH overproduction bound in \cref{fig:PS_k}).
At scales $10^{-4}\lesssim k/\text{Mpc}^{-1}\lesssim 1$, the power spectrum is constrained by the angular resolution of the current CMB measurements. However, inhomogeneities at these scales result in isotropic deviations from the usual black body spectrum and are known as CMB spectral distortions \cite{Chluba:2012we}. 
These distortions are usually categorized into two: $\mu$-distortions, associated with chemical potential that occur at early times, and Compton $y$-distortions, generated at redshifts $z \lesssim 5 \times 10^4$. A $\mu$-distortion is associated with a Bose-Einstein (BE) distribution
with $\mu \neq 0$. The most stringent constraints on spectral distortions currently come from observations made by the COBE/FIRAS experiment, which restricts $\lvert\mu\rvert \lesssim 9.0 \times 10^{-5}$ and $\lvert y \rvert \lesssim 1.5\times 10^{-5}$ at $95\%$ C.L \cite{Fixsen:1996nj}. In future, a PIXIE-like detector can investigate distortions with magnitudes $\mu\lesssim 2\times 10^{-8}$ and $y\lesssim 4 \times 10^{-9}$ \cite{A_Kogut_2011}. We also present acoustic reheating (AR) constraints on the spectrum \cite{Nakama_2014}. Note the solid lines are present and dashed are the future experiments. We find that the model parameters as shown in \cref{parmsets} satisfy the constraints of COBE/FIRAS. However, for the broad-width power spectrum, to explain the NANOGrav signal \cite{NANOGrav:2023gor, NANOGrav:2023hvm}, one not only needs to assume a diversion from the $\Lambda$CDM model but also the power spectrum conflicts with COBE/Firas. For further discussion on this issue, see \cref{appGW}.

\subsection{Primordial Black Hole formation}
The mass of the PBH (in solar mass units $M_\odot$) formation is associated with a wave vector $k$ and is given by \cite{Ballesteros:2017fsr},
\begin{align}\label{Mrange}
	M_{\text{PBH}}=3.68\left(\dfrac{\gamma_c}{0.2}\right)\left(\dfrac{g_{*}(T_f)}{106.75}\right)^{-1/6}\left(\dfrac{10^6\,\text{Mpc}^{-1}}{k}\right)^2\,M_{\odot}.
\end{align}
The fractional abundance of PBHs, $\Omega_{\text{PBH}}/\Omega_{\text{DM}}\equiv f_{\text{PBH}}$ is defined in terms of PBH mass \cite{Ballesteros:2017fsr},
\begin{align}
	\label{abund}
 f_{\text{PBH}}=\dfrac{\beta(M_{\text{PBH}})}{3.94\times 10^{-9}}\left(\dfrac{g_{*}(T_f)}{106.75}\right)^{-1/4}\left(\dfrac{\gamma_c}{0.2}\right)^{1/2}\left(\dfrac{0.12}{\Omega_{\text{DM}} h^2}\right)\left(\dfrac{M_{\text{PBH}}}{M_{\odot}}\right)^{-1/2},
\end{align}
where $h^2\Omega_{\text{DM}}=0.12$ is the current energy density of DM, $\gamma_c=0.2$ is the factor that depends on the gravitation collapse\footnote{The value of $\gamma_c$ depends on the detail of the gravitational collapse and still has uncertainties, for further details see ref.~\cite{Ando:2018qdb}.}. Using the Press Schechter approach\footnote{The fractional energy density $\beta$ can also be calculated with the Peak theory \cite{Bardeen:1985tr} 
and this approach might yield larger abundances compared to the Press Schechter approach, for detail discussion see ref.~\cite{Press:1973iz,Stamou_2023}.},  
$\beta$ the fractional energy density at the time of formation is given by \cite{Motohashi:2017kbs},
\begin{align}
\label{fracdens}
	\beta(M_{\text{PBH}})=\dfrac{1}{2\pi\sigma^2(M_{\text{PBH}})}\int_{\delta_c}^{\infty}d\delta\,\,\, \text{exp}\left(-\dfrac{\delta^2}{2\sigma^2\left(M_{\text{PBH}}\right)}\right).
\end{align}
\begin{figure}[t]
    \centering
\includegraphics[width=0.7\linewidth]{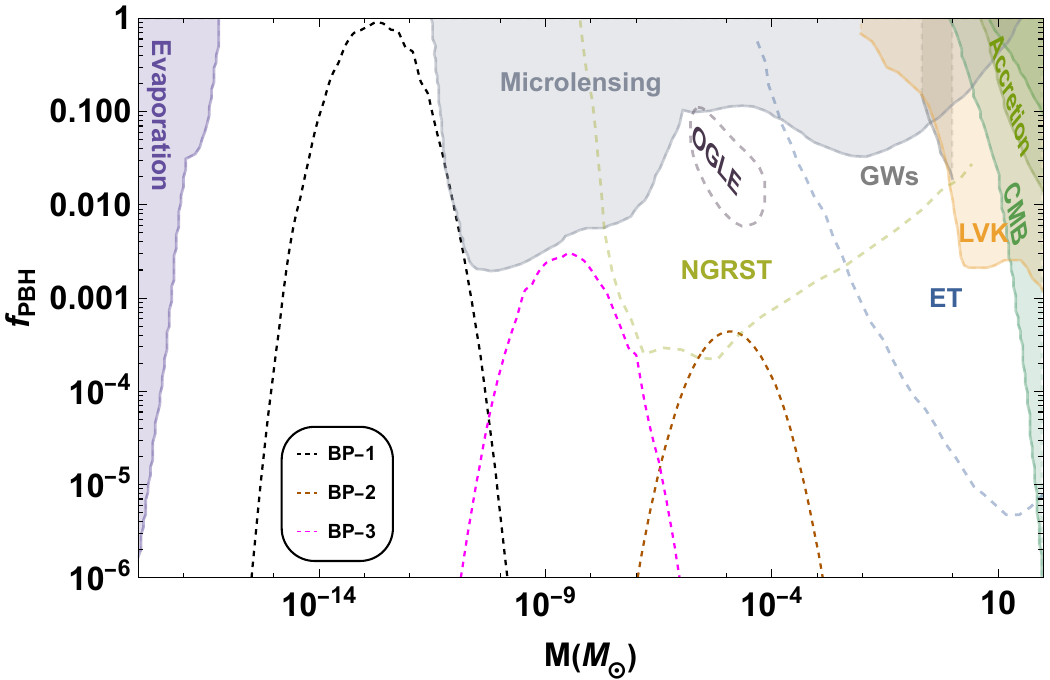}
\caption{ \it PBH abundance as DM given in \cref{abund}. The shaded regions represent the observational constraints on the PBH abundance from various experiments (solid lines present and dashed for future), see the main text for the details.}
    \label{fig:f_pbh}
\end{figure}
Here $\delta_c$ is the critical density perturbation of PBH formation, and it takes the value between $0.4-0.6$ \cite{Musco:2020jjb,Escriva:2020tak,Escriva:2019phb,Musco:2018rwt,Stamou:2023vxu} that corresponds to $\sigma^2(M_\text{PBH})\simeq10^{-2}-10^{-3}$ to explain the full abundance of DM. The variance $\sigma(M_\text{PBH})$, of the curvature perturbations, is given by
\begin{align}
	\sigma^2(M_\text{PBH}(k))=\dfrac{16}{81}\int\dfrac{dk^{'}}{k^{'}}(k^{'}/k)^4W^2(k^{'}/k) P_s(k^{'}),
\end{align} 
where $W(x)=\text{exp}(-x^2/2)$ is the Gaussian window function. The abundance of PBH \cref{abund} is shown in \cref{fig:f_pbh}. For given parameter sets in \cref{parmsets}, the PBHs explain the entire abundance of DM for BP-$1$.  However, for BP-$2$, some fraction of DM can be explained due to different observational constraints from the experiments, as shown by the shaded region in \cref{fig:f_pbh}. As from \cref{Mrange}, $M_\text{PBH}\propto k^{-2}$, therefore to explain the PBHs in PTAs the power spectrum has to be enhanced at low $k$ and large $M_\text{PBH}$ which are constrained by the Microlensing effects. Therefore, the entire abundance of DM from PBHs cannot be observed in PTAs. 

In \cref{fig:f_pbh} we depict the constraints on $f_{\rm PBH}$ (see ref.~\cite{Green:2020jor,Saha:2021pqf,Laha:2019ssq,Ray:2021mxu} for details on constraints). Evaporation of PBH via Hawking radiation sets severe constraints in: CMB\,\cite{Clark:2016nst}, EDGES\,\cite{Mittal:2021egv},  INTEGRAL\,\cite{Laha:2020ivk,Berteaud:2022tws}, Voyager\,\cite{Boudaud:2018hqb}, 511\;keV\,\cite{DeRocco:2019fjq},
EGRB\,\cite{Carr:2009jm}; HSC (Hyper-Supreme Cam)\,\cite{Niikura:2017zjd}, EROS\,\cite{EROS-2:2006ryy}, OGLE\,\cite{Niikura:2019kqi} and Icarus\,\cite{Oguri:2017ock} which are all categorized as micro-lensing related observations, several constraints arising due to modification of the CMB spectrum which happens if PBHs accrete as investigated in ref.\,\cite{Serpico:2020ehh} (see also ref.\,\cite{Piga:2022ysp}). Finally, the range around $M_{\odot}$ is constrained by LIGO-VIRGO-KAGRA (LVK) observations on PBH-PBH merger \,\cite{Franciolini:2022tfm,Kavanagh:2018ggo,Hall:2020daa,Wong:2020yig,Hutsi:2020sol,DeLuca:2021wjr,Franciolini:2021tla}), while future GW interferometer based detectors like Cosmic Explorer (CE) and Einstein Telescope (ET) are expected to set limits on the PBH abundance as shown in refs.\cite{DeLuca:2021hde,Pujolas:2021yaw,Franciolini:2022htd,Martinelli:2022elq,Franciolini:2023opt,Branchesi:2023mws}, these are shown in dashed lines in the plot. We also show future sensitivity reaches of the Nancy Grace Roman Space Telescope (NGRST) from micro-lensing, see ref. \cite{DeRocco:2023gde}.
Our prediction in \cref{parmsets}, for the BP-$1$ the PBHs explain the entire abundance of DM. While for BP-$2$ and $3$ some fraction of DM can be explained but can be tested in future experiments like NGRST. 

\medskip

\section{Scalar-induced Gravitational Waves}
\label{SIGWs}
\begin{figure}[t]
    \centering
    \includegraphics[width=0.484\linewidth,height=5.7cm]{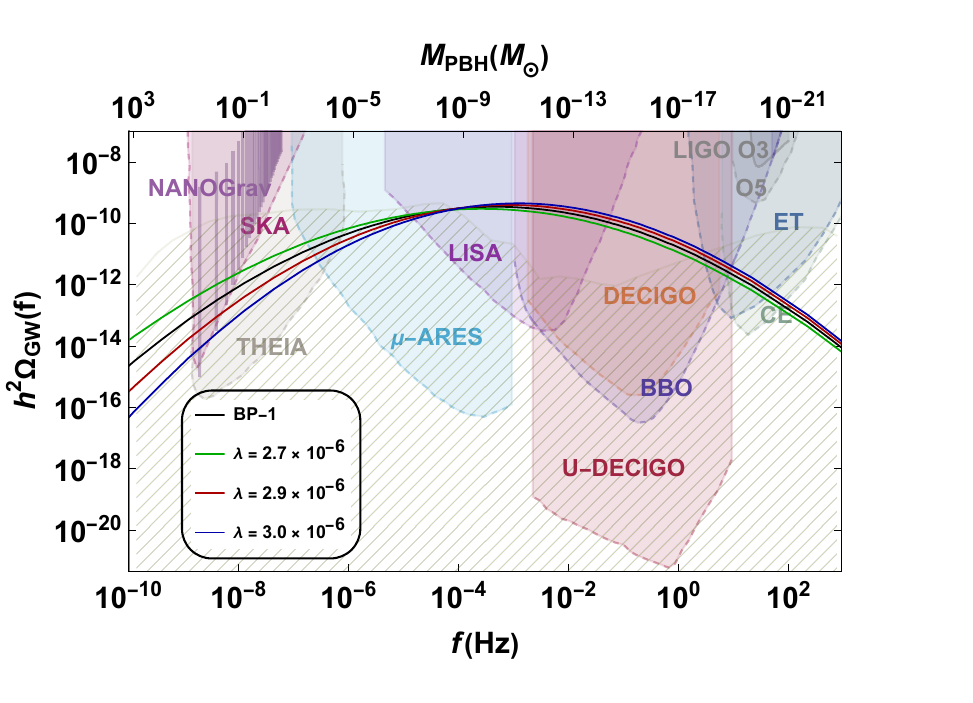}
        \quad
        \includegraphics[width=0.484\linewidth,height=5.7cm]{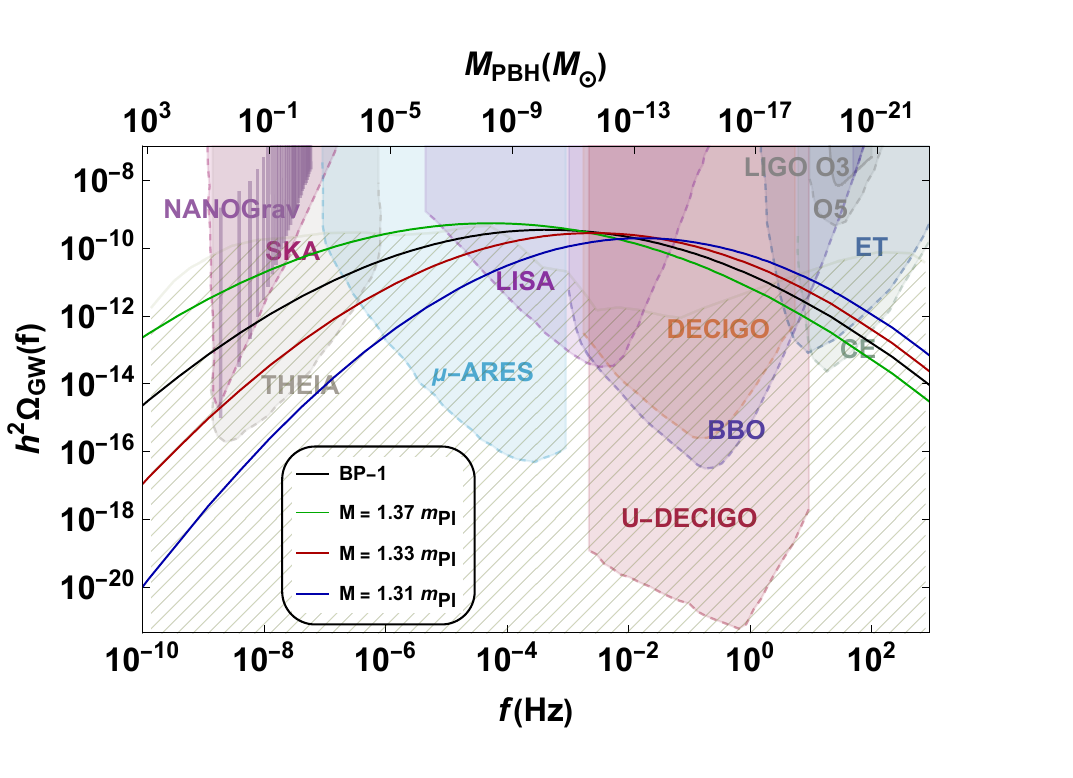}
        \quad
        \includegraphics[width=0.484\linewidth,height=5.7cm]{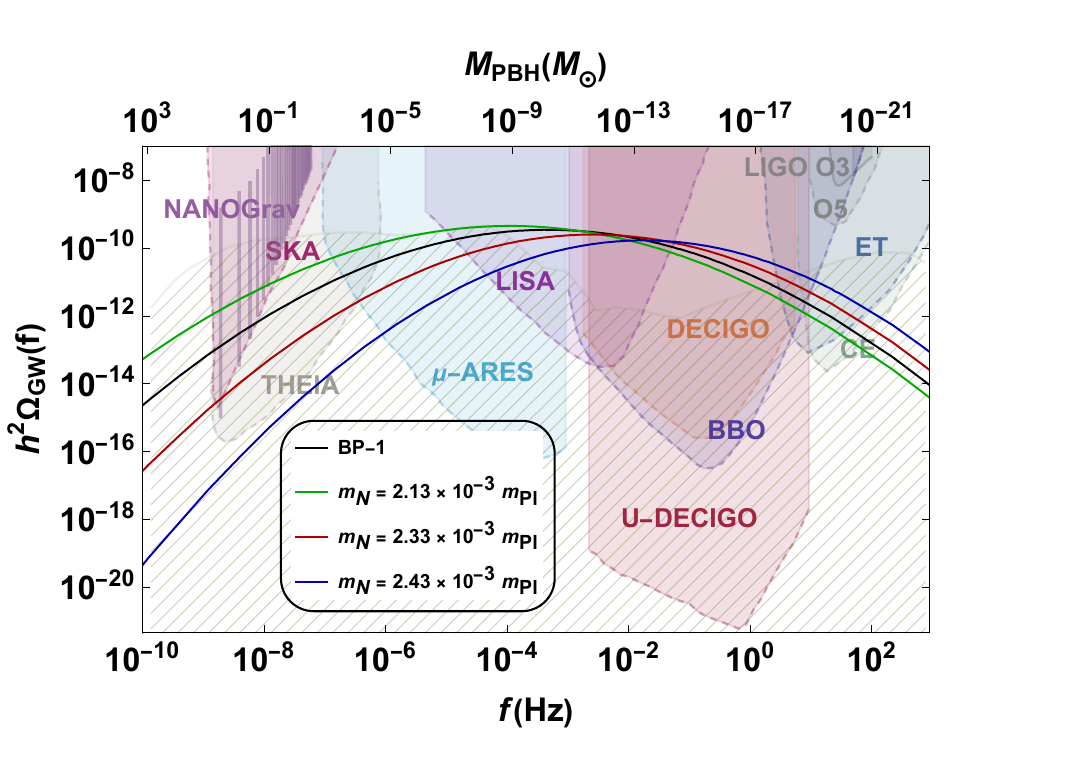}
        \quad
        \includegraphics[width=0.484\linewidth,height=5.7cm]{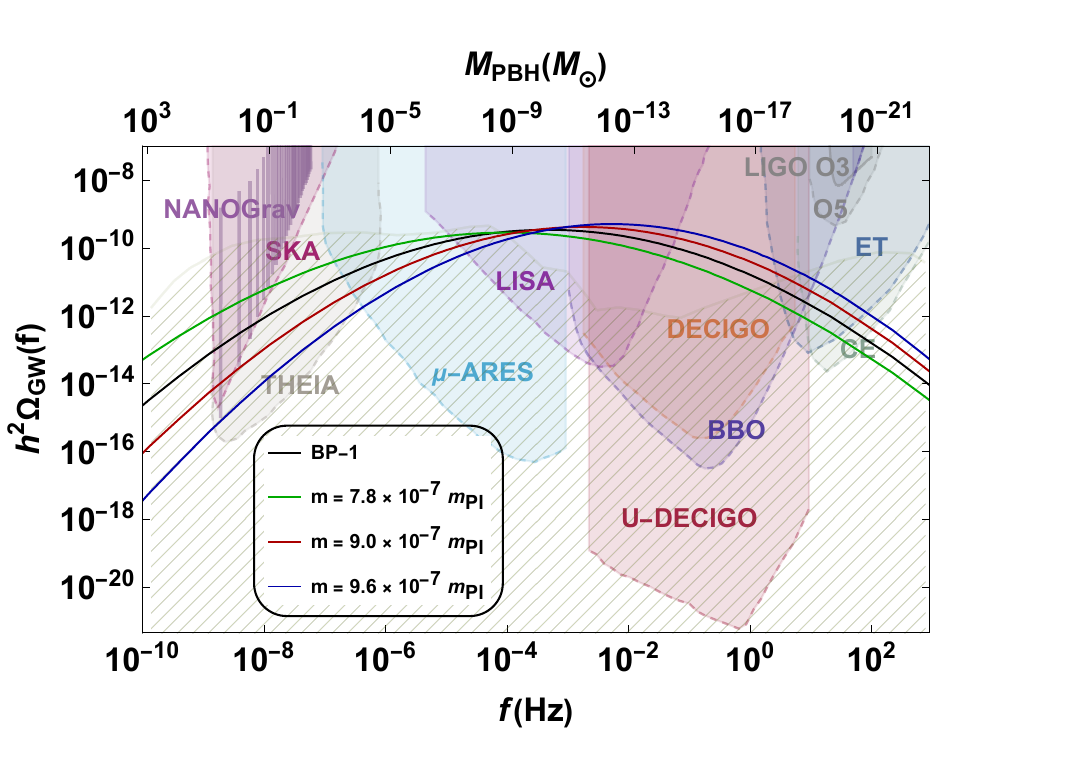}
        \quad
    \includegraphics[width=0.484\linewidth,height=5.7cm]{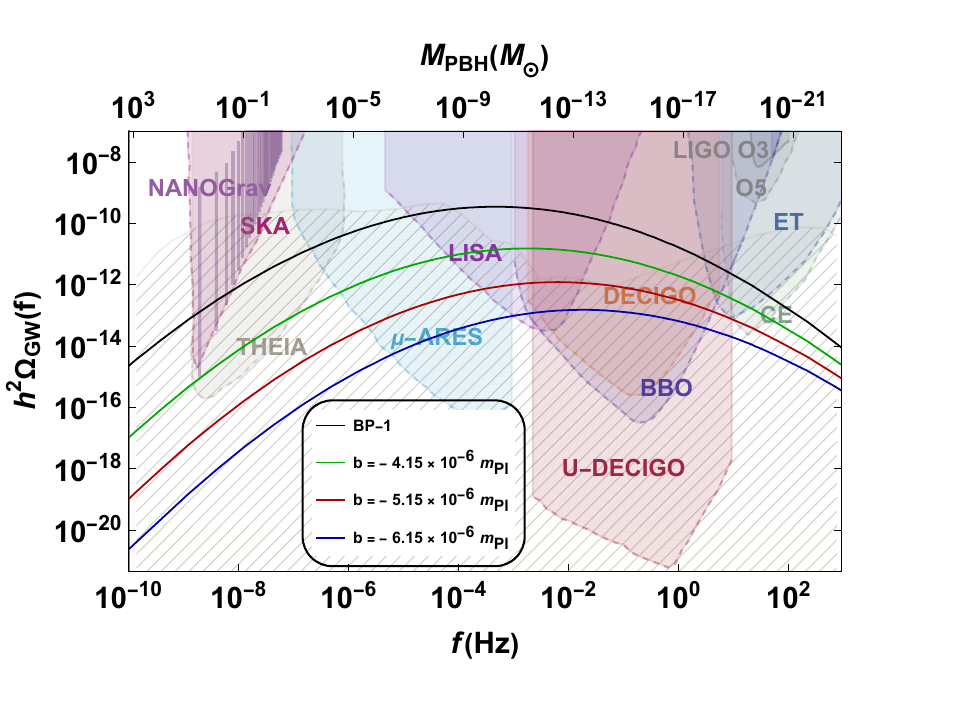}
        \quad
    \includegraphics[width=0.484\linewidth,height=5.7cm]{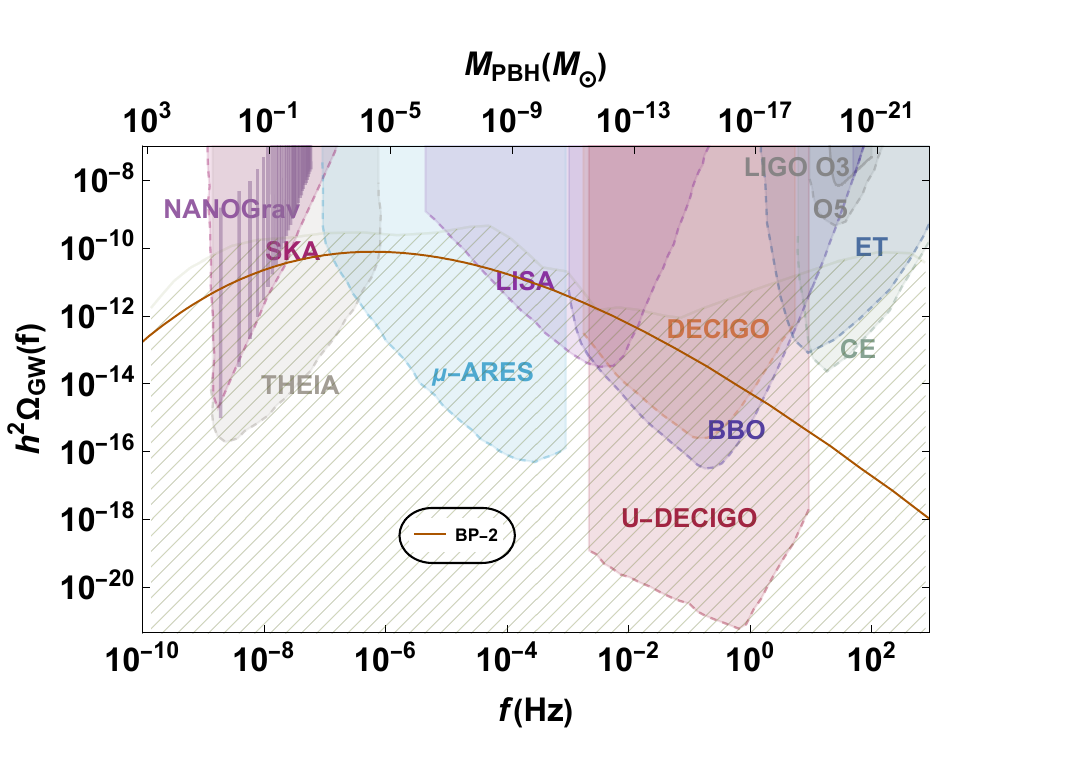}
\caption{ \it The energy density of GWs for \cref{Energdens} for the BPs given in \cref{parmsets}.  The colored shaded regions indicate the sensitivity curves of present (solid boundaries) LIGO O3 \cite{KAGRA:2021kbb}, NANOGrav \cite{NANOGrav:2023gor} and future (dashed boundaries) LIGO O5, SKA \cite{Smits:2008cf}, THEIA \cite{Garcia-Bellido:2021zgu}, LISA \cite{Baker:2019nia}, $\mu$-ARES \cite{Sesana:2019vho}, BBO \cite{Corbin:2005ny}, U-DECIGO \cite{Yagi:2011wg, Kawamura:2020pcg}, CE \cite{Reitze:2019iox} and ET \cite{Punturo:2010zz} experiments. The hatched region shows the astrophysical background 
\cite{ghoshal2023traversing}. See text for details.}
    \label{fig:Omegaf}
\end{figure}

Assuming the formation of PBHs in the radiation dominated era, the present-day energy density of the GWs in terms of scalar power spectrum \cref{Powspec}, is given by \cite{Acquaviva:2002ud, Espinosa:2018eve, DeLuca:2020agl, Maggiore:1999vm},
\begin{align}
\label{Energdens}
	\Omega_{\text{GW}}(k)&=\dfrac{c_g\,\Omega_{r}}{6}\left(\dfrac{g_{*}(T_f)}{106.75}\right)\int_{-1}^{1}dd\int_{1}^{\infty}ds\,\,P_s\left(k\dfrac{s-d}{2}\right)P_s\left(k\dfrac{s+d}{2}\right)I(d,s), \\ \notag
	I(d,s)&=\dfrac{288(d^2-1)^2(s^2-1)^2(s^2+d^2-6)^2}{(d-s)^8(d+s)^8}\left\{\left(d^2-s^2+\dfrac{d^2+s^2-6}{2}\text{ln}\left|\dfrac{s^2-3}{d^2-3}\right|\right)^2\right.\\\notag
	&\left.\dfrac{\pi^2}{4}(d^2+s^2-6)^2\Theta(s-\sqrt{3}))\right\}.
\end{align}
In the expression above, $\Omega_r=5.4\times10^{-5}$ is the present day value of energy density, $c_g=0.4$ in the SM, $\Theta$ is the Heaviside function and the effective degrees of freedom $g_{*}(T_f)$ at the temperature of PBH formation $T_f$ is $106.75$ for SM like spectrum. Furthermore, using $k=2\pi f$, $1\text{Mpc}^{-1}=0.97154\times10^{-14}\,\text{s}^{-1}$ and $h=0.68$, we transformed into the $h^2\Omega_{\text{GW}} (f)-f$ plane.
The GW spectra for the BPs in \cref{parmsets} are shown in \cref{fig:Omegaf} with the variation of different model parameters as labeled. The model can be tested in future planned experiments like SKA \cite{Smits:2008cf}, THEIA \cite{Garcia-Bellido:2021zgu}, LISA \cite{Baker:2019nia}, $\mu$-ARES \cite{Sesana:2019vho}, BBO \cite{Corbin:2005ny}, U-DECIGO \cite{Yagi:2011wg, Kawamura:2020pcg}, CE \cite{Reitze:2019iox} and ET \cite{Punturo:2010zz} experiments presented by shaded region in \cref{fig:Omegaf}. For the observation of GW spectra in the PTA band, see \cref{appGW}. 

When looking for stochastic GW background of cosmic origin we expect several astrophysical sources of GW to be present which will be like a background, mainly in the form of LIGO/VIRGO observed binary black hole (BH-BH) merging events \cite{TheLIGOScientific:2016qqj, Abbott:2016nmj, TheLIGOScientific:2016pea,Abbott:2017vtc, Abbott:2017gyy, Abbott:2017oio, LIGOScientific:2018mvr} and binary neutron star (NS-NS) events \cite{TheLIGOScientific:2017qsa,Venumadhav:2019lyq}.
Therefore to distinguish scalar-induced GWs of cosmic origin, the NS and BH foreground may be subtracted with sensitivities of BBO and ET or CE windows, particularly in the range $\Omega_{\rm GW} \simeq 10^{-15}$ \cite{Cutler:2005qq} and $\Omega_{\rm GW} \simeq 10^{-13}$ \cite{Regimbau:2016ike}.
In the LISA window, the binary white dwarf galactic and extra-galactic (WD-WD) may be of greater significance than the NS-NS and BH-BH foregrounds \cite{Farmer:2003pa, Rosado:2011kv, Moore:2014lga} and can be subtracted \cite{Kosenko:1998mv} with the expected sensitivity at $\Omega_{\rm GW} \simeq 10^{-13}$ \cite{Adams:2010vc, Adams:2013qma}. This subtraction procedure alongside the fact that the GW spectrum generated by the astrophysical foreground increased with frequency as $f^{2/3}$ \cite{Zhu:2012xw} is different from the GW spectrum generated by second-order GWs from radiative hybrid inflation ( at low frequency $f^{3/2}$, at higher frequency $f^{-3/2}$). Now, one will be able to pin down the GW signals from the scalar-induced gravitational wave source from the hybrid inflation as we predict.

\medskip


\subsection*{Signal-to-noise Ratio}

\begin{figure}[t]
    \centering
    \includegraphics[width=0.484\linewidth,height=5.0cm]{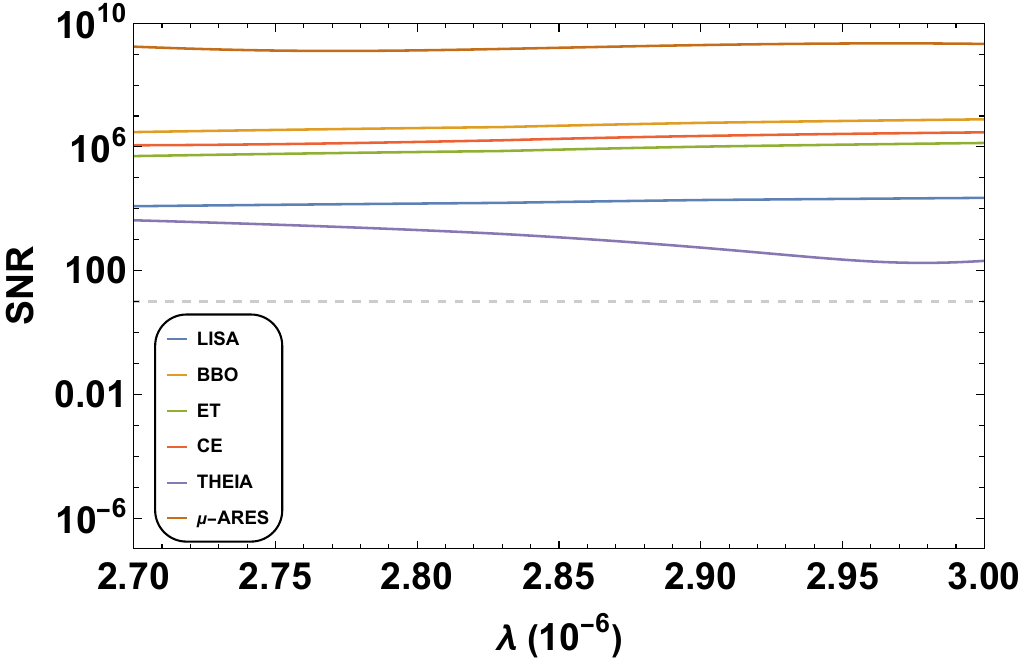}
    \quad
        \includegraphics[width=0.484\linewidth,height=5.0cm]{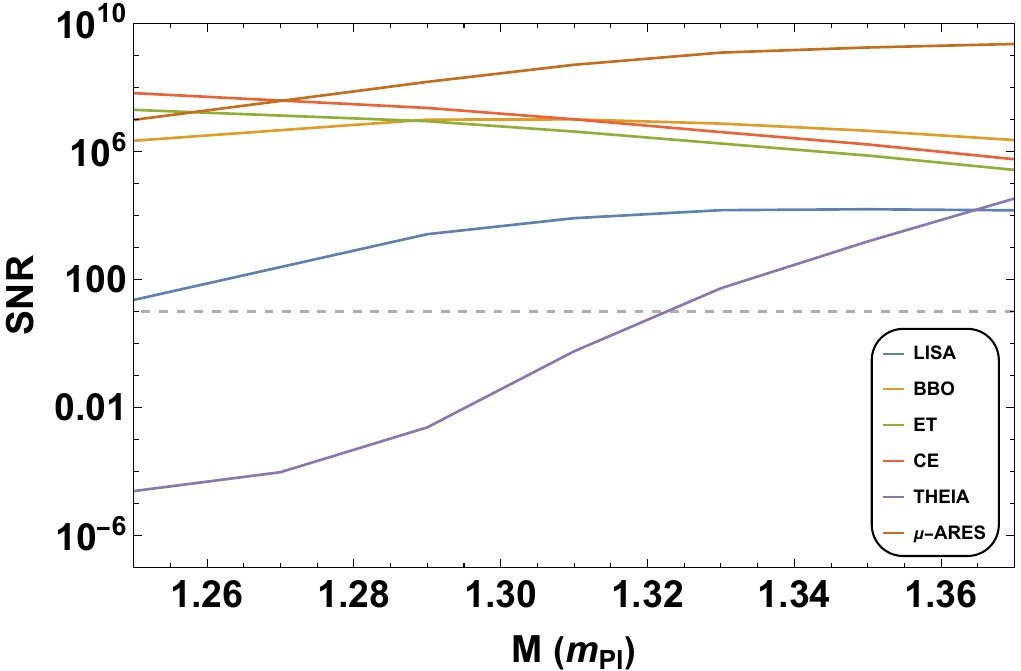}
        \quad
        \includegraphics[width=0.484\linewidth,height=5.0cm]{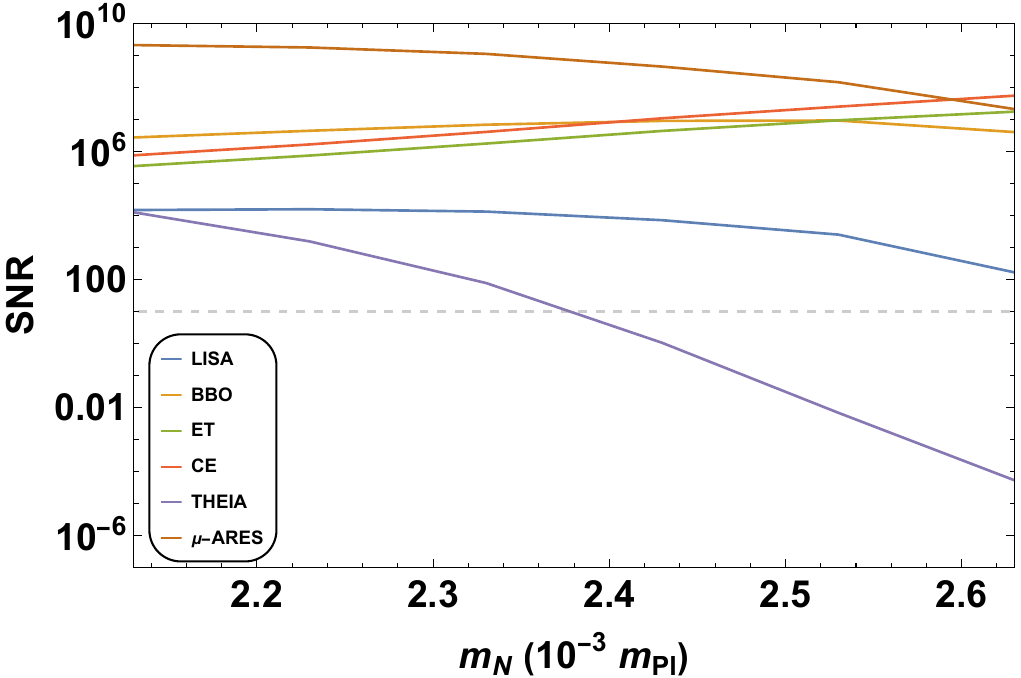}
             \quad
        \includegraphics[width=0.484\linewidth,height=5.0cm]{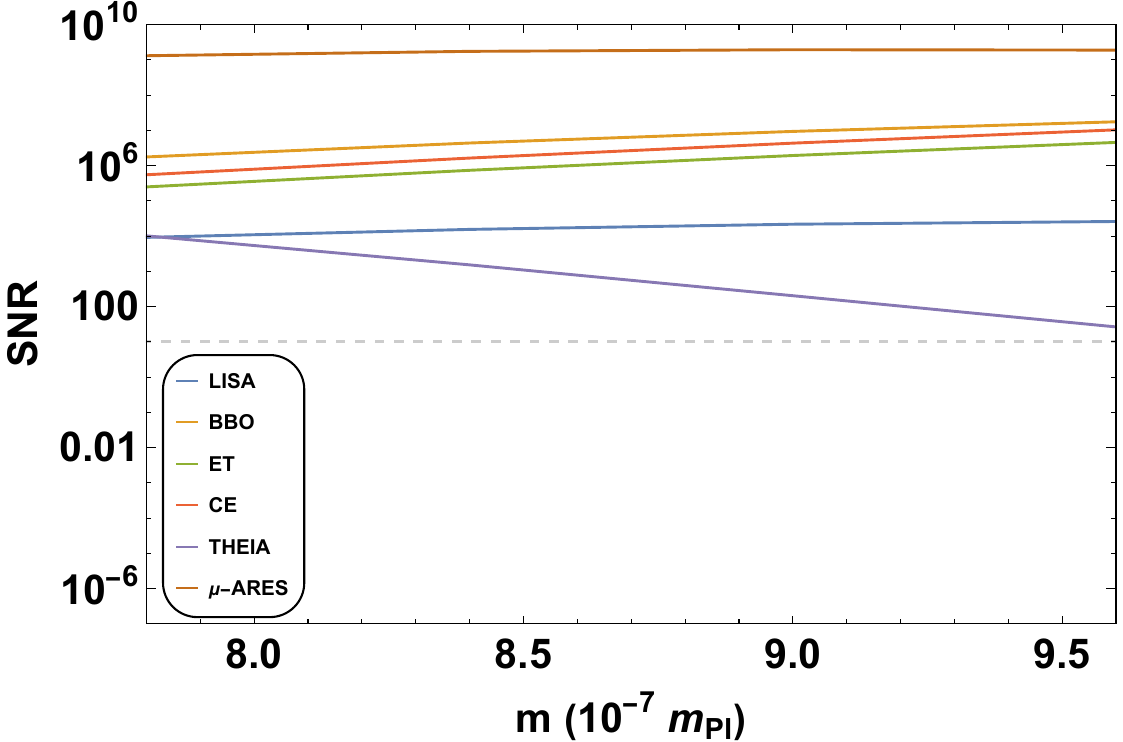}
       \quad \includegraphics[width=0.484\linewidth,height=5.0cm]{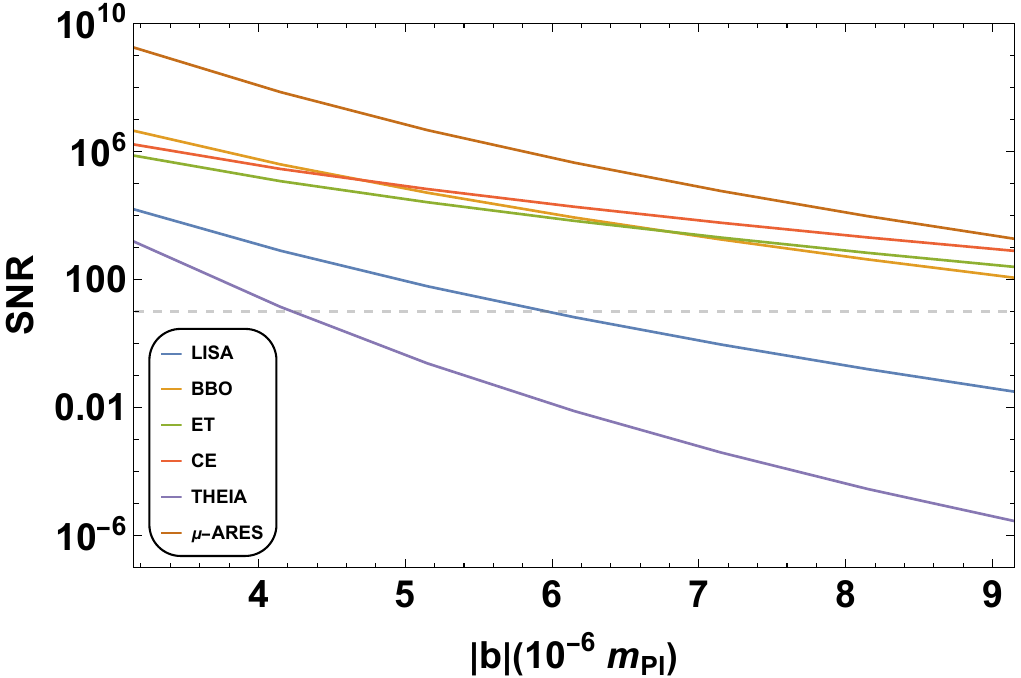}
\caption{ \it Signal-to-noise ratio (SNR) as a function of various model parameters indicating projections of future sensitivity of different experiments as indicated by the color coding in the inset. The dashed gray line indicates SNR $=10$.}
    \label{fig:SNR}
\end{figure}

The ability of an interferometer to quantify SGWB signal with an energy density $\Omega_\text{GW}(f)$ for observation time $T_\text{obs}$ is quantified as the signal-to-noise ratio (SNR). 
To estimate the SNR, we refer to the standard way of computation \cite{Caprini:2018mtu},
\begin{equation} 
\label{eq:SNR}
	\text{SNR} \equiv  \left[ T_\text{obs} \int_{f_\text{min}}^{f_\text{max}} \left( \frac{\Omega_{\text{GW}}(f)}{\Omega_{\text{Noise}}(f)}  \right)^2  d {f}
	\right]^{1/2} \, ,
\end{equation}
where $f_\text{min}$, $f_\text{max}$ denotes the detector bandwidth. We have utilized the noise curve $\Omega_{\text{Noise}}(f)$ for a given experiment and have assumed the duration of each GW mission to be $T_\text{obs}=4$~years. We present the SNR as a function of different model parameters in \cref{fig:SNR} and show SNR=$10$ as a reference with a gray dashed line. We find that most of the future planned experiments LISA, BBO, DECIGO, ET, CE, THEIA and $\mu$-ARES are all capable of testing the model with SNR~$ > 10$.

\begin{figure}[t]
    \centering
         \includegraphics[width=0.484\linewidth,height=5.0cm]{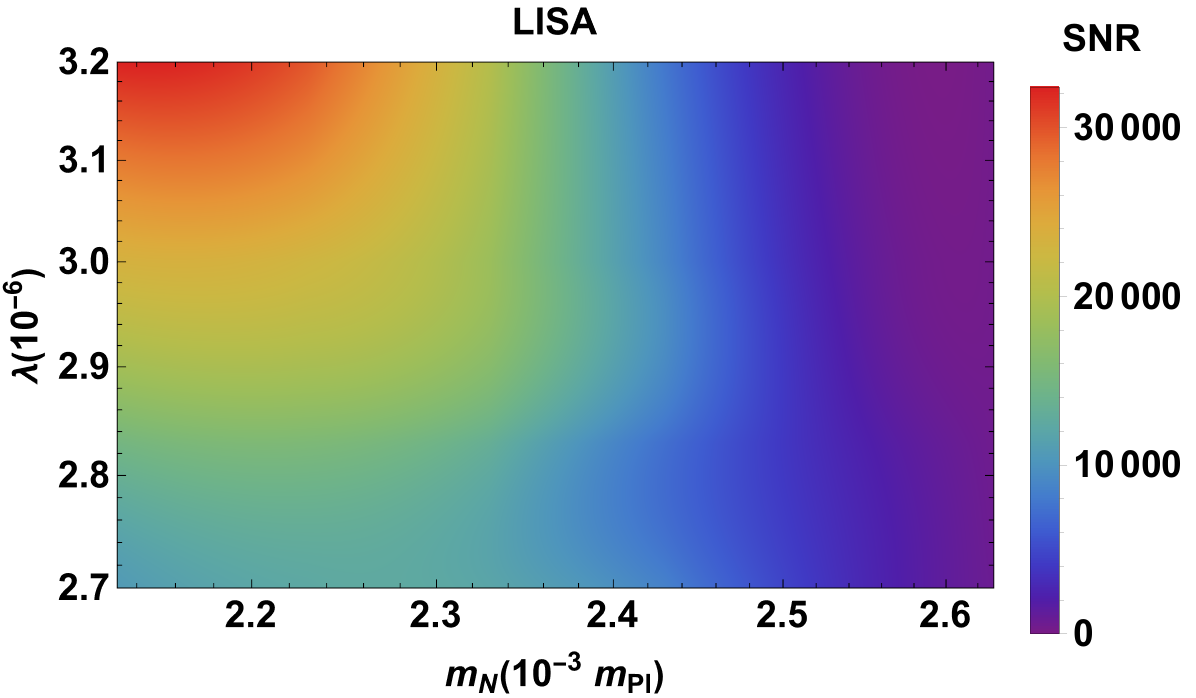}
        \quad
        \includegraphics[width=0.484\linewidth,height=5.0cm]{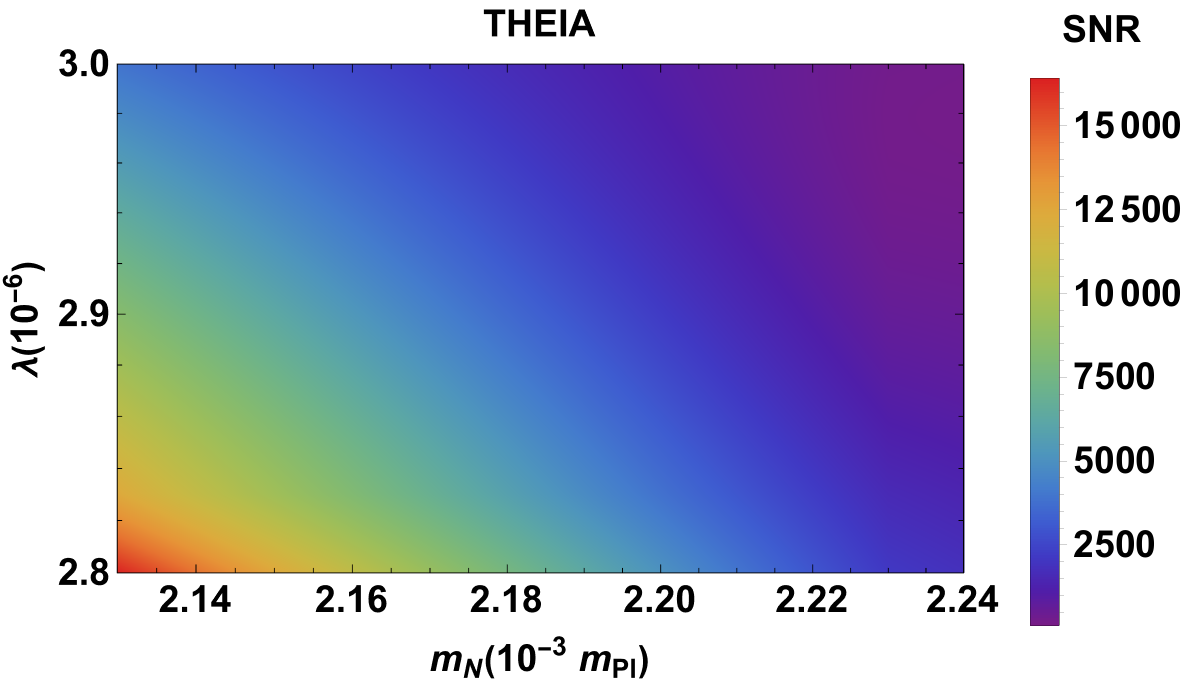}
        \quad
        \includegraphics[width=0.484\linewidth,height=5.0cm]{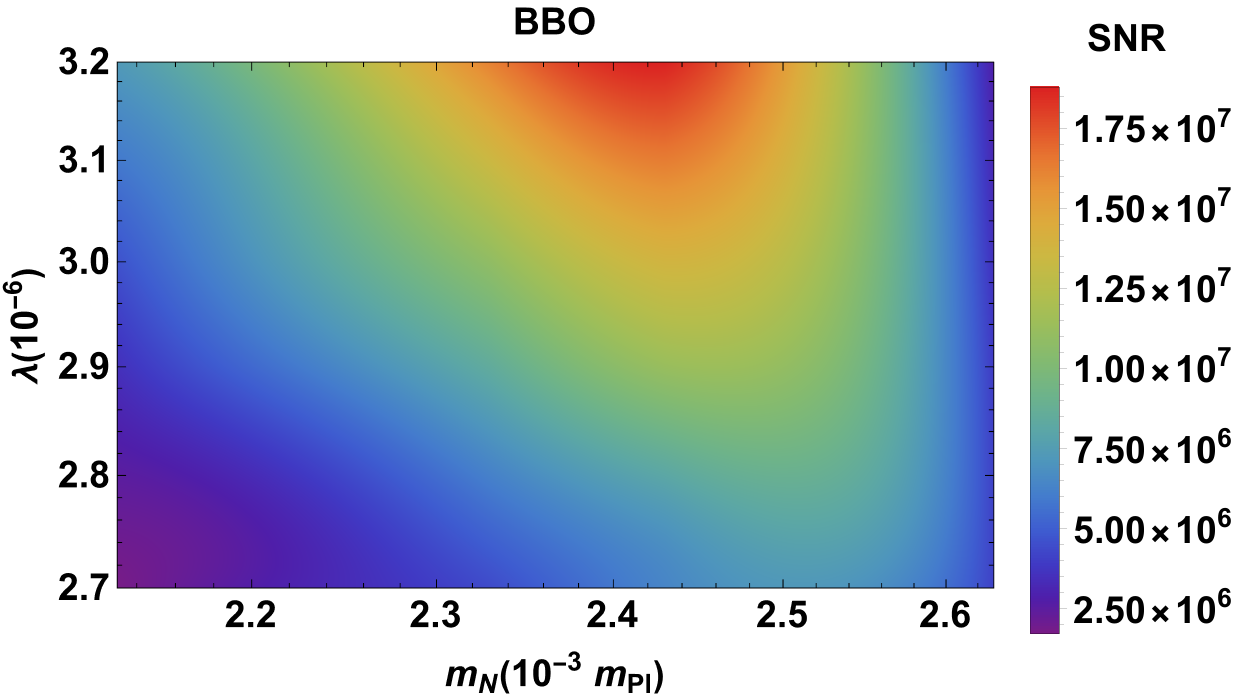}
         \quad
        \includegraphics[width=0.484\linewidth,height=5.0cm]{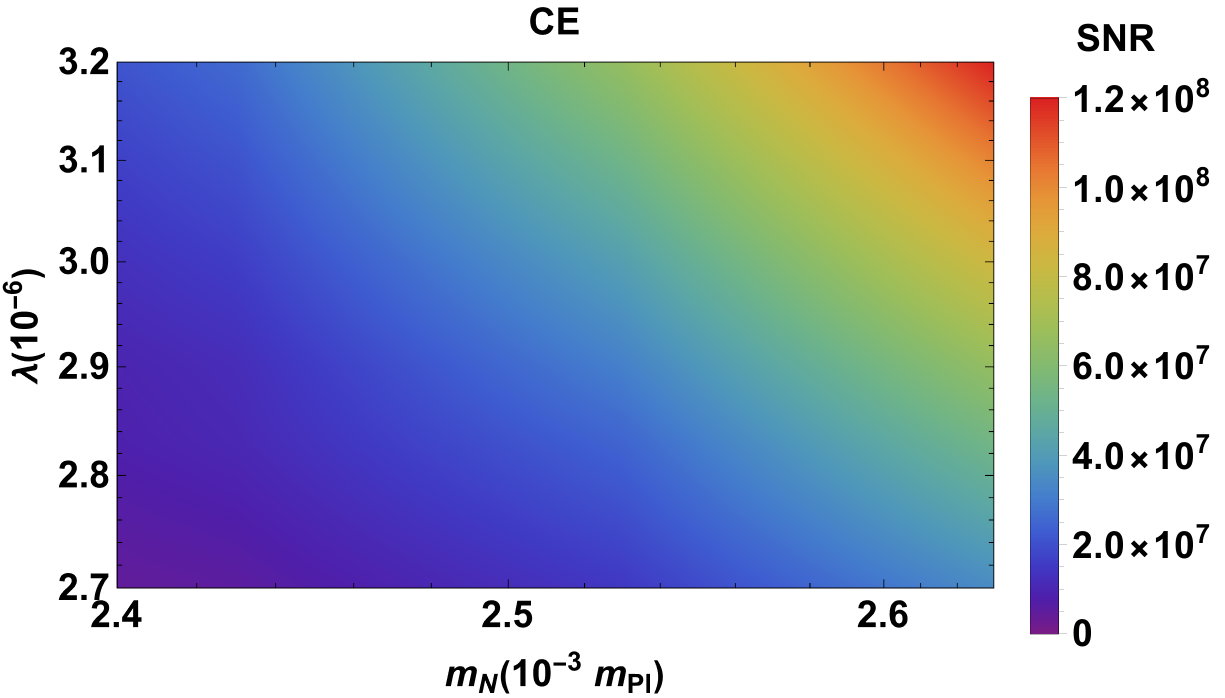}
        \quad
        \includegraphics[width=0.484\linewidth,height=5.0cm]{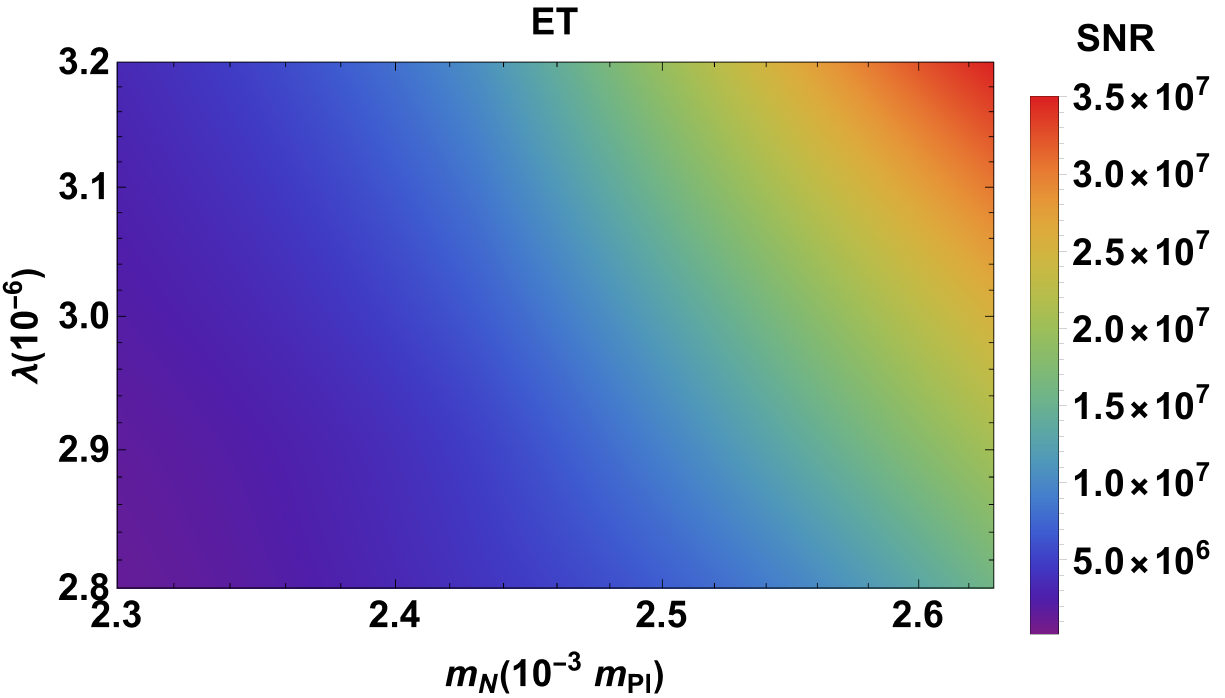}
        \quad
        \includegraphics[width=0.484\linewidth,height=5.0cm]{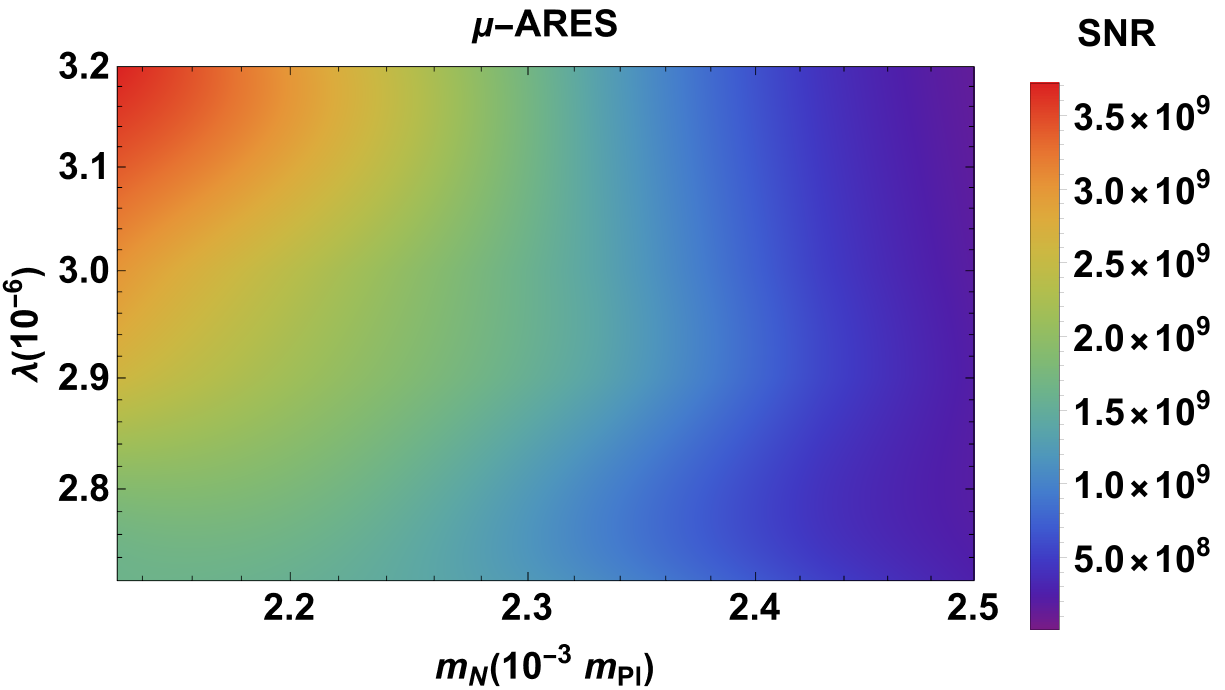}
\caption{ \it Parameter space in the $\lambda$ vs. $m_N$ with varying SNR are shown by the vertical bar legends for LISA, THEIA, BBO, CE, ET and $\mu$-ARES as labeled.}
    \label{fig:ParamSNRapp}
\end{figure}

The behavior of SNR can be comprehended from \cref{fig:Omegaf} where we vary one parameter and the rest of them are fixed according to BP-$1$ in \cref{parmsets}. Variations of $\lambda$ in the top left plot in \cref{fig:Omegaf} do not have a large impact on the parameter space within the sensitivity curves like LISA, BBO, ET, CE and $\mu$-ARES, this gives a slight increase in SNR by increasing $\lambda$. There is a noticeable decrease in THEIA, since, by increasing $\lambda$, the power spectrum will span less space of the sensitivity curve for THEIA, as shown in the top left plot of \cref{fig:SNR}. Variation of $M$ in the top right of \cref{fig:Omegaf} predicts less sensitivity for ET, CE, BBO and more sensitivity for LISA $\mu$-ARES, THEIA with increasing $M$. This behavior is shown in the upper left panel of \cref{fig:SNR}. In the middle left of \cref{fig:Omegaf}, one can see increasing sensitivity for ET and CE and a decrease for LISA, BBO, THEIA and $\mu$-ARES with increasing $m_N$. This is consistent with the SNR shown in the middle left of \cref{fig:SNR}. Parameter $m$ affects the power spectrum in such a way that if it leaves the sensitivity curve at a lower frequency, it will span a slightly more parameter space at a higher frequency, therefore increasing SNR with increasing $m$ is depicted in the middle right of \cref{fig:SNR} except for THEIA. Increasing the coefficient of the linear term $b$, reduces the peak of the spectrum and therefore SNR will decrease as shown in the third row of \cref{fig:SNR}. 

\begin{figure}[t]
    \centering
        \includegraphics[width=0.484\linewidth,height=5.0cm]{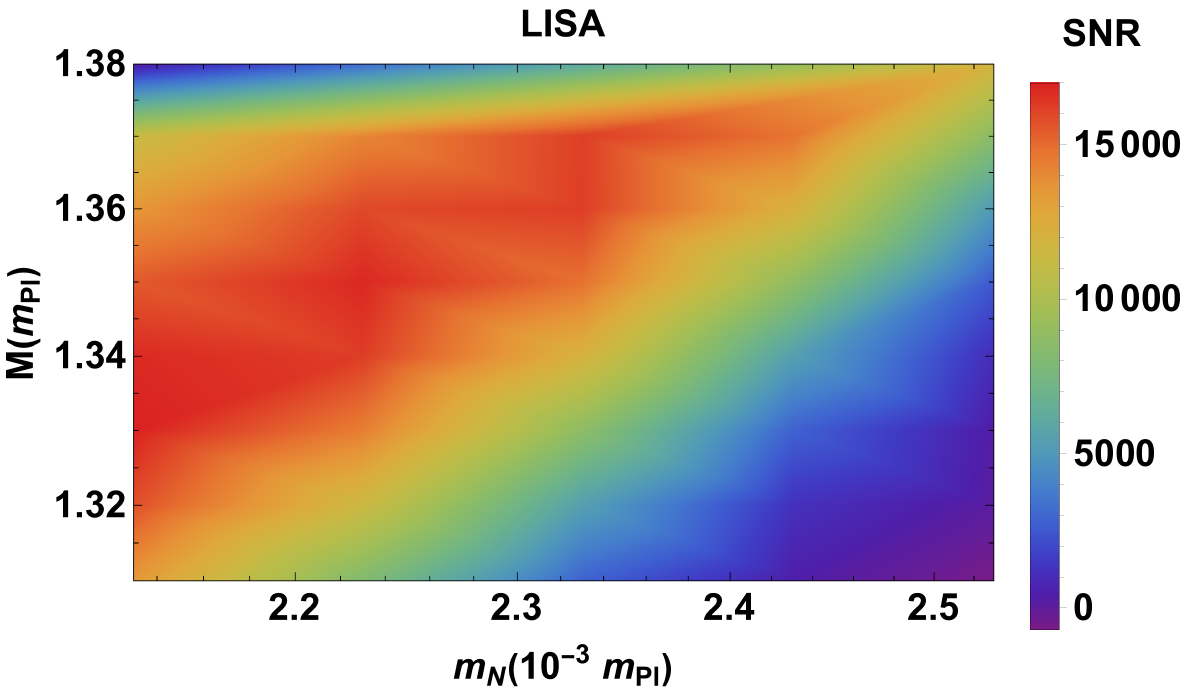}
          \quad
        \includegraphics[width=0.484\linewidth,height=5.0cm]{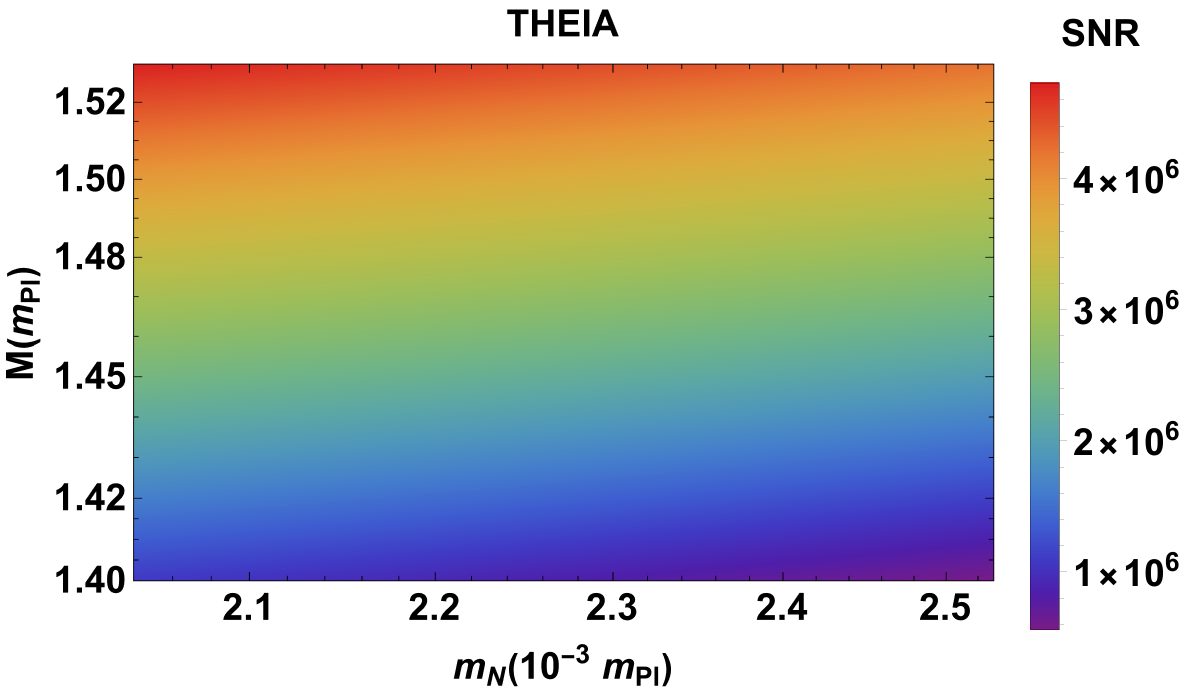}
         \quad   \includegraphics[width=0.484\linewidth,height=5.0cm]{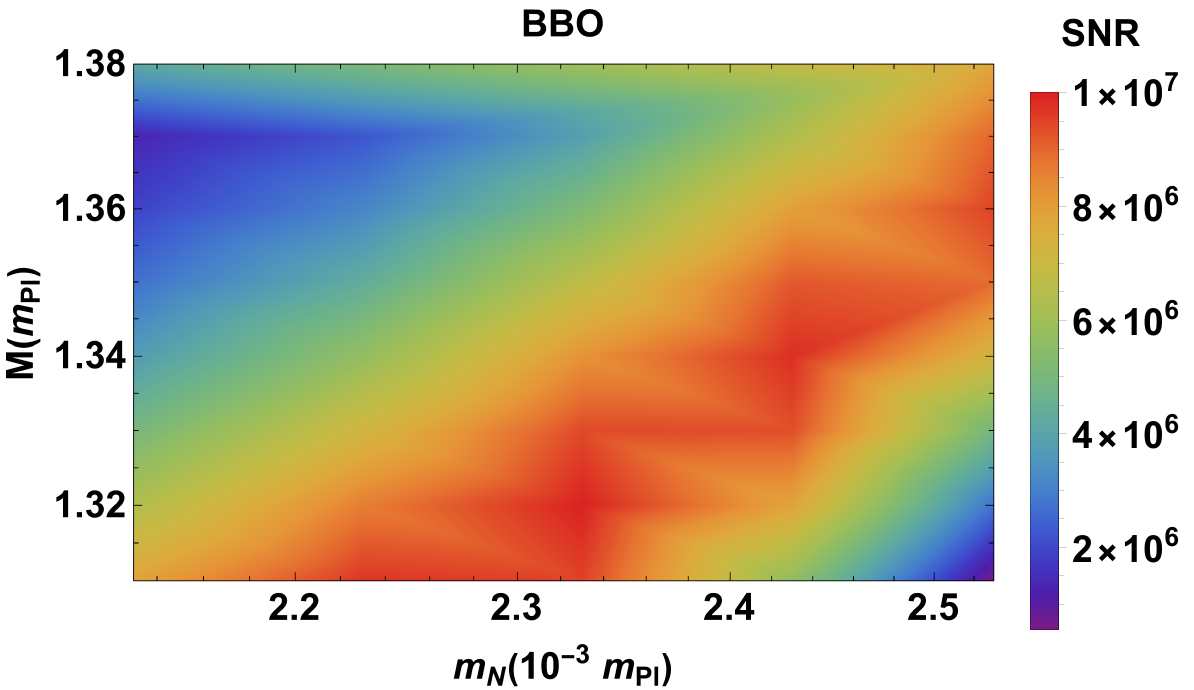}
    \quad
        \includegraphics[width=0.484\linewidth,height=5.0cm]{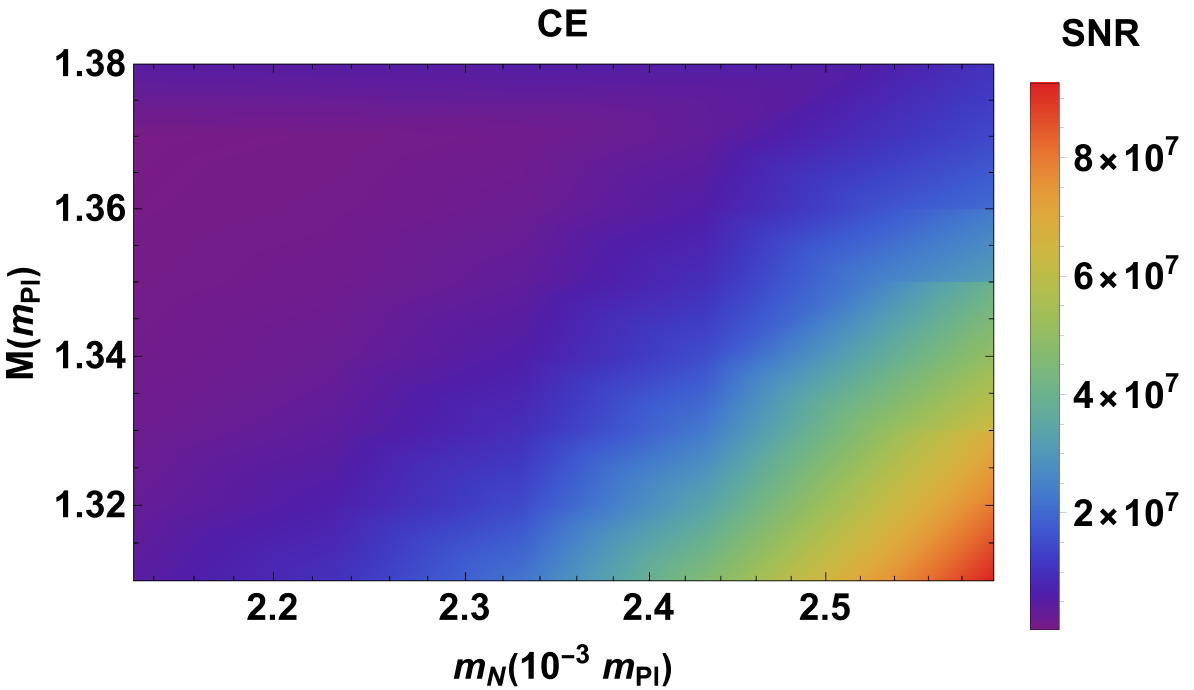}
        \quad
        \includegraphics[width=0.484\linewidth,height=5.0cm]{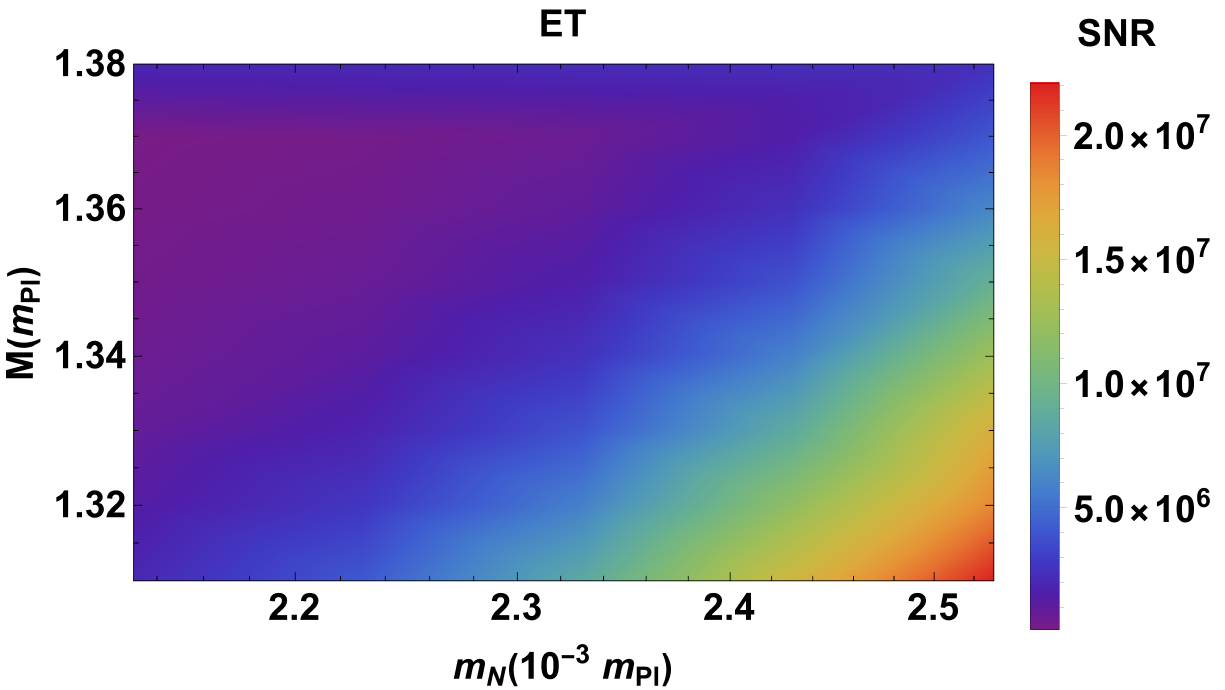}
        \quad
        \includegraphics[width=0.484\linewidth,height=5.0cm]{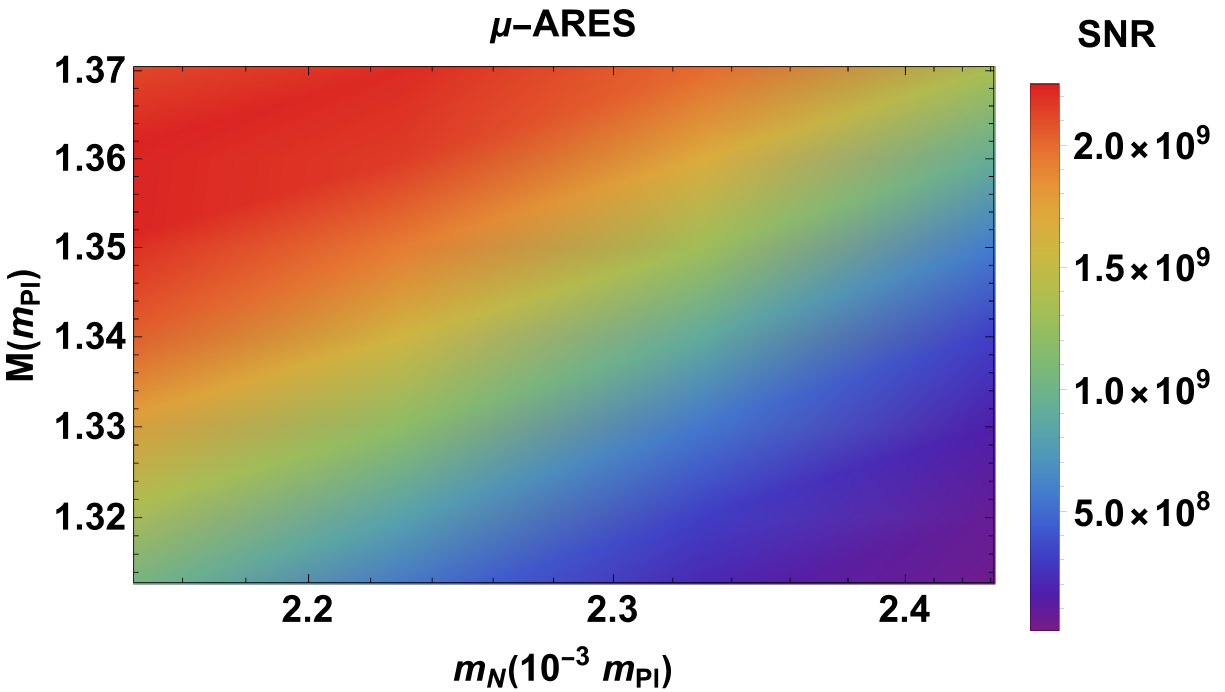}
\caption{ \it Parameter space in the $M$ vs. $m_N$ with varying SNR are shown by the vertical bar legends for LISA, THEIA, BBO, CE, ET and $\mu$-ARES as labeled.}
    \label{fig:ParamSNR}
\end{figure}

In \cref{fig:ParamSNRapp}, we present the correlation between the parameter space for the model parameters $\lambda$ vs. $m_N$ with the variation in SNR shown by a vertical bar for LISA, THEIA, BBO, CE, ET and $\mu$-ARES as indicated. In \cref{fig:ParamSNRapp}, the SNR for LISA and $\mu$-ARES increases with the decreasing $m_N$, and increasing $\lambda$, this feature is depicted in the top left and bottom right of \cref{fig:ParamSNRapp}. For ET and CE, increasing both $m_N$ and $\lambda$ increases the SNR; see the top right and bottom left of \cref{fig:ParamSNRapp}. For THEIA, the SNR increases with decreasing $m_N$ and $\lambda$, see the upper right of \cref{fig:ParamSNRapp}. For BBO, the behavior is different due to its sensitivity frequency range, SNR will increase with increasing $\lambda$ for a particular range of $m_N$, see the middle right of \cref{fig:ParamSNRapp}.
The correlation parameter space for $M$ vs. $m_N$ for the indicated experiments is given in \cref{fig:ParamSNR}. For LISA and BBO the SNR increases for some intermediate values of $M$ and $m_N$, for the rest of the parameter space, it decreases. For THEIA, SNR increases with increasing $M$ and for $\mu$-ARES the increasing $M$ and decreasing $m_N$ gives the larger SNR. For ET and CE, the behavior is similar, small $M$ and large $m_N$ increase the SNR; see \cref{fig:ParamSNR}.

\medskip
\section{Fine-tuning Estimates}
\label{finetuneapp}
The single-field inflationary models, where the PBHs arise due to inflection points, etc. require a high level of fine-tuning of the parameters involved in the enhancement of the power spectrum at small scales \cite{Ballesteros:2017fsr,Ghoshal:2023wri}. In this section, we show that, unlike single-field inflation, the amount of fine-tuning is much smaller in hybrid inflation. To estimate the required amount of fine-tuning, we calculate a quantity $\Delta_x$ given by
\begin{align}
    \Delta_x = \text{Max}\bigg|\dfrac{\partial \ln P_s^\text{Peak}}{\partial \ln x}\bigg|,
\end{align}
where $x\in\{m,M,m_N,\lambda\}$ is the model parameter. Evaluating numerically, a fine-tuning estimate for theory parameters is given in \cref{finetune}. The larger the $\Delta_x$ is, the larger the amount of required fine-tuning. Note that we did not consider the case for $b$, since the variation of $b$ does not keep the height of the power spectrum fixed and therefore affects the abundance of DM. 
\begin{table}[tbh!]
	  \fontsize{12pt}{13pt}
	\caption{ \it Fine-tuning (FT) estimate of model parameters for BP$-1$ in \cref{parmsets} with a peak of the spectrum around $5\times 10^{-3}$.  } 
	\centering 
	\begin{adjustbox}{max width=\columnwidth}
		\begin{tabular}{|M{1.5cm} |M{1cm} |M{1cm} |M{1cm} |M{1cm}|M{1cm}|M{1cm}|M{1cm}|M{1cm}|M{1cm}|M{1cm}|M{1cm}|}
			\hline
   \bf{$\Delta_x$}& \bf{$\Delta_m$}&\bf{$\Delta_M$}&\bf{$\Delta_{m_N}$}&\bf{$\Delta_\lambda$}
			\\ [0.6ex] 
			\hline\hline
			$\text{FT}$ &$2$  &$15$  & $5$ & $ 2 $\\
			\hline 
		\end{tabular}
	\end{adjustbox}
	\label{finetune}
\end{table}
The maximum fine-tuning we obtain is $15$, which is almost five orders of magnitude smaller than single-field inflation \cite{Stamou:2021qdk} and one order of magnitude smaller than supersymmetric hybrid inflation \cite{Spanos:2021hpk}\footnote{Ref.~\cite{Spanos:2021hpk} estimated fine tuning of some of parameterisation of the model while we present explicitly the fine-tuning of all parameters.}.

\begin{figure}[H]
    \centering
    \includegraphics[width=0.47\linewidth,height=4.4cm]{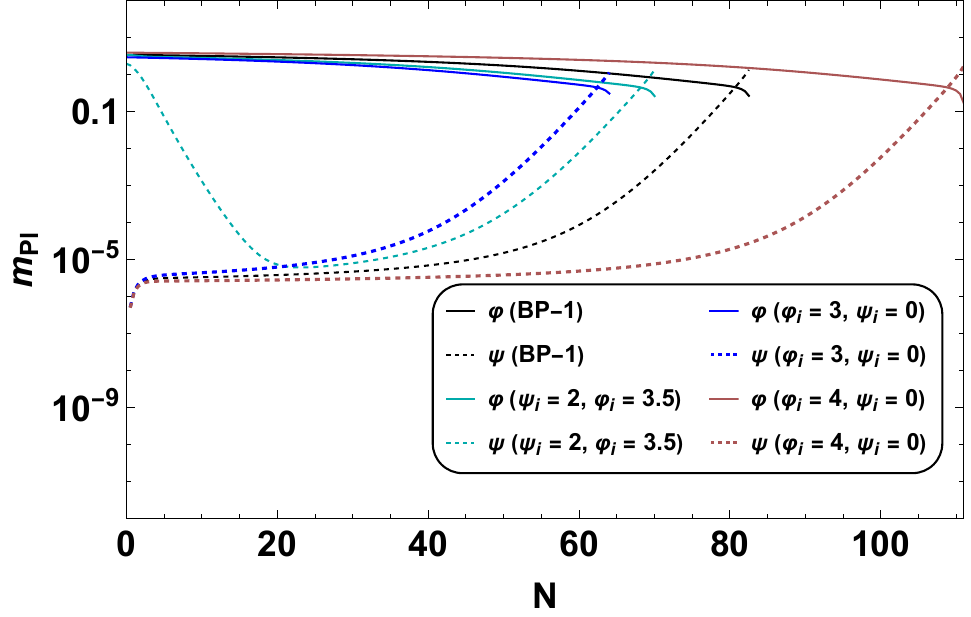}
    \quad
        \includegraphics[width=0.47\linewidth,height=4.4cm]{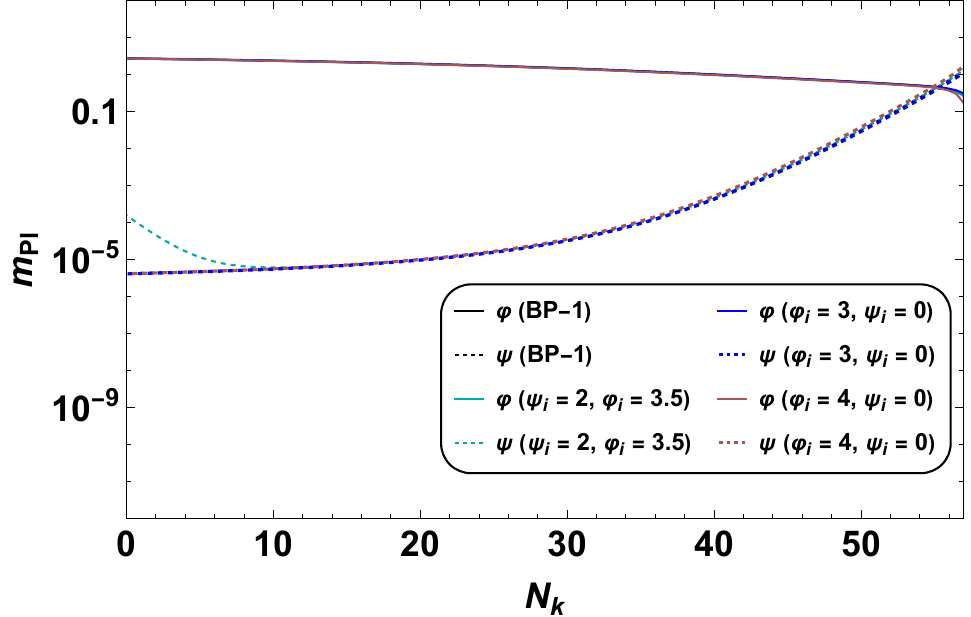}
\caption{ \it In the left panel, variation of the field values with respect to the number of e-folds is shown for different initial conditions. In the right panel, we have shown the field value from the horizon exit till the end of inflation. }
    \label{fig:psiphiapp}
\end{figure}

As can be seen in the right panel of \cref{fig:psiphiapp}, variation of initial conditions has no impact within our constrained parameter space. Due to attractor behavior, the power spectrum is not affected by the initial condition of $\phi$ within a constrained number of e-folds. In the right panel of \cref{fig:psiphiapp}, we show the field evolution for the total number of e-folds for different initial conditions. In the left panel of \cref{fig:psiphiapp}, we present the last $57$ e-folds from horizon exit till the end of inflation, contributing as a BP-$1$. Therefore, the initial conditions do not contribute to the fine-tuning estimate within the constrained number of e-folds. For further details on the initial conditions see \cite{Braglia:2022phb}.
Also note that due to the presence of the linear term in \cref{canonpoten}, $\psi$ will not relax exactly at zero but is slightly displaced depending on the sign of the coefficient $b$. Therefore, even if $\psi_i=0$, the field will evolve with the number of e-folds and can be seen in \cref{fig:psiphiapp}.
\section{Reheating}
\label{reheat}
\begin{figure}[H]
    \centering
    \includegraphics[width=0.6\linewidth]{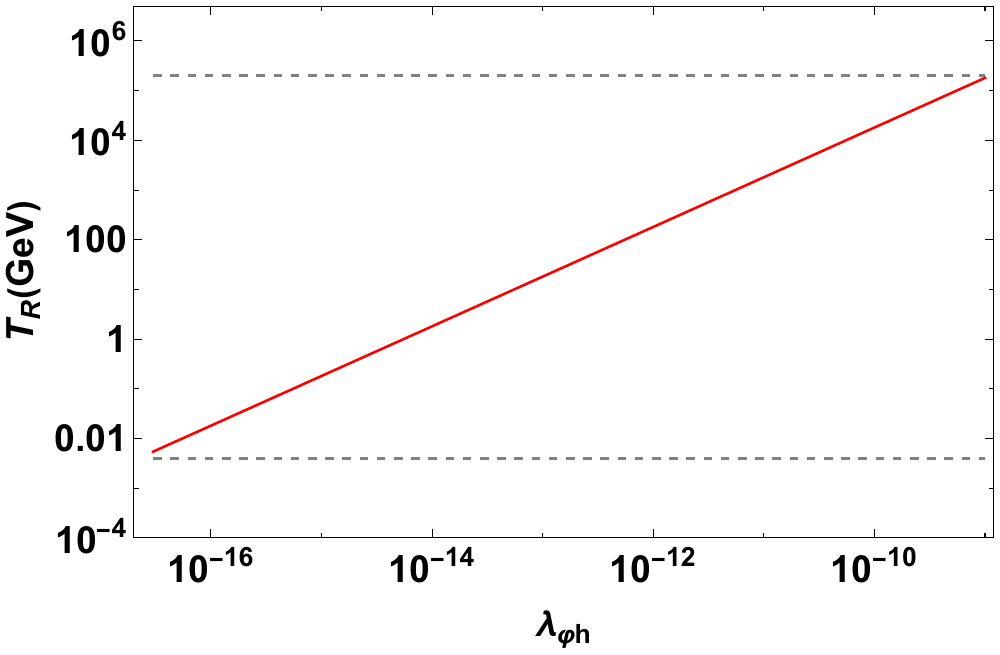}
    \caption{ \it Reheat temperature as a function of $\lambda_{\varphi h}$ is shown in red color and the gray dashed lines are the upper and lower bounds, see the main text for an explanation.}
    \label{fig:reheat}
\end{figure}

Let us now consider the SM Higgs $h$ and its coupling with $\varphi$ and $\psi$ that is conducive to reheating the universe. The potential is written as,

\begin{align} \label{higgsterms}
    V=\lambda_h\left(h^2-\dfrac{v_h^2}{2}\right)^2+2\,\lambda_{\phi\,h}\,\left(h^2-\dfrac{v_h^2}{2}\right)^2\phi^2+2\,\lambda_{\psi h}\left(\psi^2-\dfrac{M^2}{2}\right)\left(h^2-\dfrac{v_h^2}{2}\right),
\end{align}
where $v_h$ is the vacuum expectation value (VEV) of the SM Higgs, $\lambda_{\phi h}$, $\lambda_h$, $\lambda_{\psi h}$ are dimensionless couplings.
For simplicity, we assume $ {\lambda_{\phi h},\lambda_{\psi h}}$ to be very tiny such that the inflation and waterfall transition are not affected by the SM Higgs dynamics. However, $\lambda\gg \lambda_{\phi h}$ need not be exactly zero but small, as this will be responsible for the reheating\footnote{A detailed analysis involving RGE of the three fields has been studied in ref.~\cite{Ibrahim:2022cqs}. In this context, our focus is solely on upholding consistency in accordance with the findings presented therein.}. Now, once inflation ends, the inflaton 
will decay and reheat the universe. The inflaton predominantly decays into the SM Higgs and the corresponding reheat temperature is estimated assuming perturbative reheating by,
\begin{align}
T_R\simeq\sqrt{\left(\dfrac{90}{\pi^2\,g_\star}\right)^{1/2}\Gamma_\varphi\,m_\text{Pl}}.
\end{align}
Here, $g_\star$ is the effective degrees of freedom and $\Gamma_\varphi(\varphi\rightarrow hh)$ is the decay rate of $\varphi$ given by,
\begin{align}
    \Gamma_\varphi\simeq\dfrac{\lambda_{\varphi h}^2}{8\,\pi}m.
\end{align}
The coupling $\lambda_{\varphi h}$ is bounded below from the reheat temperature, that is $T_R\gtrsim 4$ MeV and above from the assumption $\lambda \gg \lambda_{\varphi h}$ therefore, $10^{-17}\lesssim \lambda_{\varphi h} \lesssim 10^{-9}$ and is shown in \cref{fig:reheat}. The maximum reheat temperature we acquire is $2\times 10^5$ GeV. We see that even very tiny $\lambda_{\varphi h}$ leads to successful reheating of the universe and this is also consistent with our earlier assumption. Such choices of the very small Higgs-portal coupling allowed for SM Higgs dynamics not to let the SM Higgs acquire too large quantum fluctuation during inflation as explicitly studied in ref.~\cite{Ibrahim:2022cqs}\footnote{ For PBHs in nonminimal derivative coupling inflation with quartic potential and reheating we refer the reader to~\cite{Heydari:2021gea}.}.  For inflaton-Higgs coupling considered in our analysis $\lambda_{\varphi h} \ll 10^{-10}$, non-perturbative preheating particle production is negligible, therefore we consider simple perturbative reheating estimates from inflaton decay \cite{Kofman:1997yn, Lozanov:2019jxc}.
\section*{Right-handed Neutrino, Seesaw and Non-thermal Leptogenesis}
\label{RHN}
In this section, we speculate a possibility that our proposed dark fermion could be the heavy right-handed neutrino (RHN) which via seesaw generates tiny masses for the SM neutrinos. Considering the relevant portion of Lagrangian, which is responsible for reheating and seesaw, 
\begin{align}
\label{lagseesaw}
     \mathscr{L}\supset-\dfrac{1}{2}y_h\,h\,\Bar{v}_L\,N-\dfrac{1}{2}y\,\phi\,\Bar{N}\,N-\dfrac{1}{2}Y\,\psi\,\Bar{N}\,N-\dfrac{1}{2}m_N\,\Bar{N}\,N.
\end{align}
 SM Higgs $h$, gives mass to the neutrino through Yukawa
coupling $y_h$, with the SM neutrino $v_L$. Accordingly, the neutrinos acquire the mass via the seesaw mechanism \cite{Croon:2019dfw}, 
\begin{align}
\label{neutmass}
    m_v = \dfrac{y_h^2\,v_h^2}{m_N},
\end{align}
where $v_h = 174$ GeV is the  VEV of the Higgs field.

Consider the BP-$3$ in \cref{parmsets}, the effective mass of the waterfall field \cref{waterfallmass} at the end of inflation is $M_{\psi}=3.3\times 10^{-4}\, m_\text{Pl}> 2\,m_N$. This kinematic bound has to be satisfied for the successful decay of the waterfall field into the RHNs. The waterfall field after the end of inflation will decay into RHNs and the corresponding decay rate is given by, 
\begin{align}
    \Gamma_\psi(\psi\rightarrow NN)\simeq\dfrac{Y^2}{8\,\pi}M_\psi.
\end{align}
The reheating temperature is given by,
\begin{align}
\label{reheatneut}
    T_R\simeq\sqrt{\left(\dfrac{90}{\pi^2\,g_\star}\right)^{1/2}\Gamma_\psi\,m_\text{Pl}}.
\end{align}
 The decay channel $\Gamma_\psi$ is important for successful leptogenesis. Taking into account the non-thermal leptogenesis and assuming instantaneous decay of the waterfall field into RHN and subsequent decays of RHN to SM particles to generate the lepton asymmetry and transfer it to baryon asymmetry via sphaleron \cite{Ghoshal:2022fud},
\begin{align}
    \dfrac{n_B}{s}\simeq 4.43\times 10^{-17}\,\text{Br}_{\psi\rightarrow NN}\left(\dfrac{T_R}{1\,\text{GeV}}\right)\left(\dfrac{2\,m_N}{M_\psi}\right)\left(\dfrac{m_{\nu_3}}{0.05\,\text{eV}}\right) \delta_\text{eff}.
\end{align}
Here, $m_{\nu_3}\simeq 0.05$ eV and $\delta_\text{eff} \leq 1 $ is the CP violating phase factor. The branching ratio Br~$=1$ if the dominant decay of the waterfall field is into RHNs.
 For the BP-$3$ in \cref{parmsets} and for observed baryon to entropy ratio; $n_B/s\simeq8.7\times 10^{-11}$ gives the reheat temperature,
\begin{align}
T_R\simeq 5 \times 10^9\,\, \text{GeV},
\end{align}
to explain the entire baryon asymmetry of the universe via the leptogenesis mechanism.
\begin{figure}[t]
    \centering
    \includegraphics[width=0.471\linewidth,height=5.7cm]{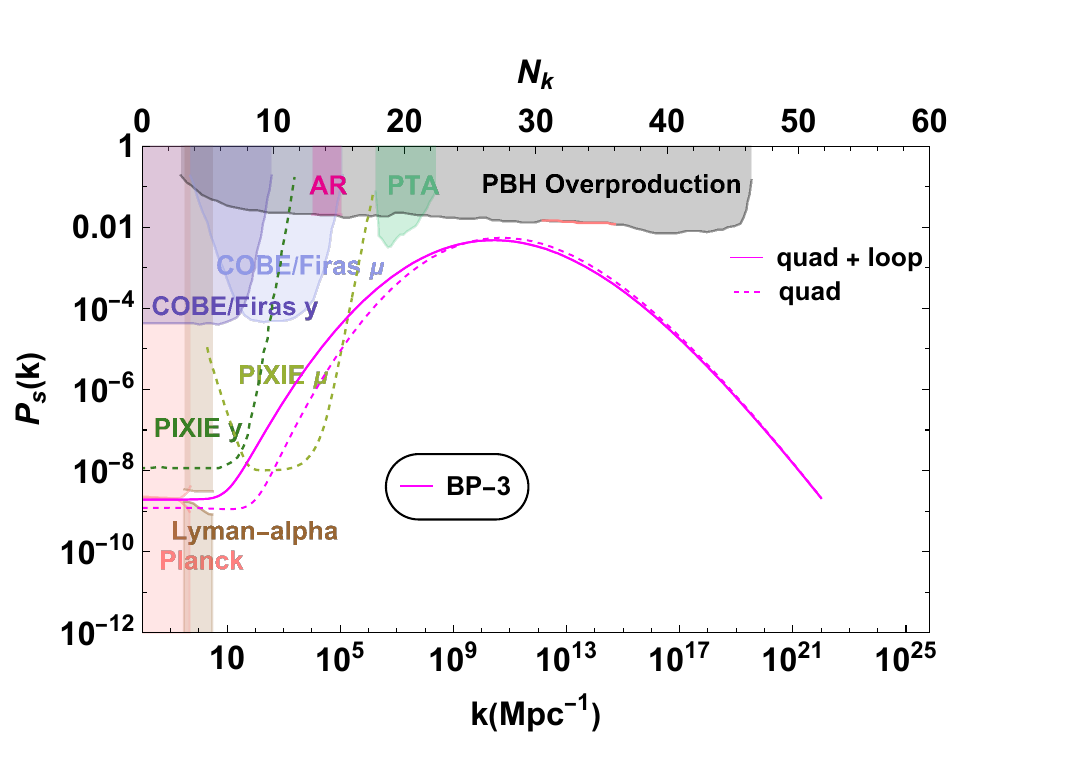}
    \quad
    \includegraphics[width=0.471\linewidth,height=5.7cm]{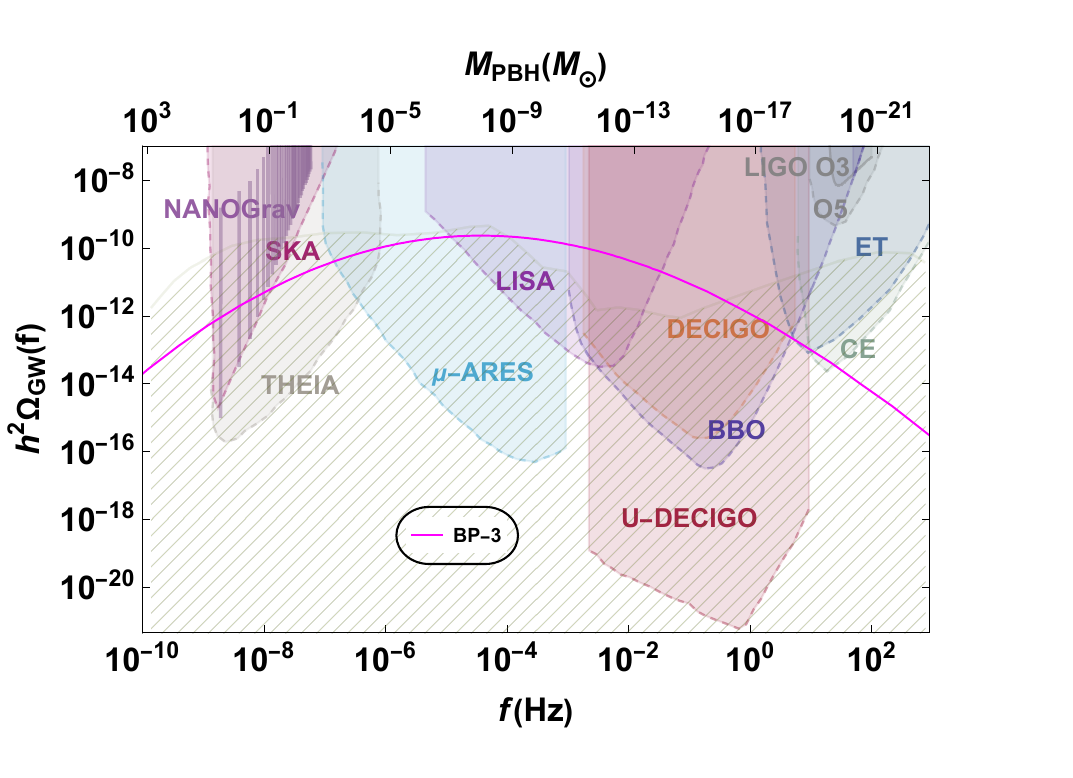}
    \caption{ \it Scalar power spectrum (left) and GW power spectrum (right) for BP-$3$ where the dark fermion is regarded as the RHN.}
    \label{fig:GWneutrino}
\end{figure}
For BP-$3$, $m_N \gg T_R$ which ensures that RHN is not produced from the thermal radiation bath so the only contribution to RHN abundance comes from waterfall field decay, therefore the non-thermal nature of leptogenesis\footnote{The maximum temperature of the universe $T_\text{max}$ is always a few orders larger than the reheat temperature \cite{Giudice:2000ex, Garcia:2020eof}. Therefore, $m_N>T_\text{max}$ is the more realistic condition than $m_N \gg T_R$ for nonthermalization of the RHN neutrino \cite{Datta:2022jic}. For detailed analysis on $T_\text{max}$, see \cite{Kolb:2003ke}.}.
The impact of the presence of RHN on the inflaton and waterfall field evolution modifying the critical waterfall transition point with respect to the number of e-folds is shown in \cref{fig:phipsi} (see BP-3). Due to this modification, the predictions in the scalar and GW spectra get affected as shown in \cref{fig:GWneutrino}. PBHs here may explain some fraction of DM, see \cref{fig:f_pbh} and is testable in the upcoming Nancy Grace Roman Space Telescope.

\section*{Higgs Vacuum Stability and Quantum Fluctuations}

 In this section, we briefly discuss the fate of the universe due to the presence of the Higgs field during inflation. The electroweak vacuum stability or meta-stability demands that the Hubble rate $H$ during inflation should be less than the Higgs instability scale $\Lambda$ that lies between $10^9\lesssim \Lambda/\text{GeV} \lesssim 10^{12}$ depending upon the precise measurement of the top quark mass \cite{Elias-Miro:2011sqh, Datta:2013mta, Branchina:2014usa}. 
 The Higgs field may get pushed over the barrier due to quantum fluctuations that destabilize the Higgs field during inflation if the typical momentum $k\sim H$ is greater than the potential barrier leading to decay of the electroweak vacuum \cite{Gross:2015bea, Fumagalli:2019ohr, Bezrukov:2019ylq}. Usually, this problem is avoided by the presence of new physics at the instability scale $\Lambda$ or via Higgs-inflaton couplings \cite{Gross:2015bea,Lebedev:2021xey, Lebedev:2012sy,Ema:2017ckf}. For our BPs in \cref{parmsets}, $H \sim 10^{12}$ GeV we just give an example where the Higgs vacuum is still not destabilized however a careful analysis involving renormalization group equation (RGEs) following ref. \cite{Ibrahim:2022cqs} could be needed to be studied which we plan to take up in future publication.

The effective mass $m_h$ of the Higgs field during inflation should be larger than $H$ in order to ensure that the SM Higgs does not affect the hybrid inflationary scenario we illustrated. For this purpose, following ref.~\cite{Ibrahim:2022cqs} one may modify the potential \cref{hpoten}, by introducing the SM singlet $S$ with the following interactions terms,
\begin{align}
\label{vacuuminst}
    V(S,h)\simeq \left(-2\,\lambda_{\phi S}\,\phi^2 + \lambda_S\left(S^2-\dfrac{v_S^2}{2}\right)+2\,\lambda_{\psi S}\left(\psi^2-\dfrac{M^2}{2}\right)+2\,\lambda_{hS}\left(h^2-\dfrac{v_h^2}{2}\right) \right)\left(S^2-\dfrac{v_S^2}{2}\right).
\end{align}
 where it can be assumed that the couplings $\lambda_{hS} \ll \lambda$ and $\lambda_{\phi S}$, $\lambda_{\psi S}, \lambda_{S}$ are negligibly small such that they do not affect the inflaton and waterfall dynamics. Due to the additional $S$ field, $v_S$ can be tuned such that the SM Higgs effective mass can be large during inflation such that $m_h \gg H$ condition is satisfied and the Higgs vacuum remains stable \cite{Ibrahim:2022cqs}. Therefore, the Higgs fluctuations do not excite iso-curvature perturbations and the Higgs field is frozen during inflation for the Hubble rate around $10^{12}$ GeV for the BPs we considered in \cref{parmsets} at the end of inflation. 
 
 To explain the baryon asymmetry of the universe via leptogenesis, the heavy neutrino $m_N \gtrsim 10^9$ GeV \cite{Ipek:2018sai,Ma:2006ci}. Moreover, to keep the model perturbative involving Yukawas for seesaw, there is an upper bound on the RHN that is $m_N\lesssim 10^{15}$ GeV see \cite{Ipek:2018sai,Berbig:2023yyy} which is ensured in our choice in BP-$3$ as shown in \cref{parmsets}. Lastly, Yukawa coupling $y_h$ being $O(1)$ may affect the RGEs of the SM Higgs and make the vacuum unstable, which we also ensure in our choice of BP-$3$ in \cref{parmsets}. For detailed analysis involving RGE, we refer the reader to ref. \cite{Ipek:2018sai}\footnote{In fact, if the SM Higgs becomes unstable during inflation it could give rise to interesting features in the form of PBHs, scalar-induced GW and non- 
  Gaussianities due to the tachyonic behavior studied in refs.~\cite{Espinosa:2017sgp,Espinosa:2018eve,Espinosa:2018euj,Gross:2018ivp,Shakya:2023zvs}.}.

\medskip
\section{Discussion and Conclusion}
\label{concl}


In summary, we presented a two-field inflationary model based on the original hybrid model and studied the generation of both gravitational waves and primordial black holes.
 The effective scalar potential derived by hybrid models has the advantage that they do not require a high level of fine-tuning of the parameters to describe an amplification
in the scalar power spectrum. As the issue of fine-tuning is regarded as a major problematic feature in many proposed inflationary models, studying hybrid models should be a plausible scenario to describe enhancement in the scalar power spectrum. To predict acceptable values for the spectral index $n_s$ we introduced $\alpha$-attractor specific scenario involving a pole in the kinetic term following \cite{Braglia:2022phb} where we evaluated the prediction for the cosmological constraints. In our proposed model we have achieved acceptable values for both spectral index ($n_s\simeq0.964$) and tensor-to-scalar ratio ($r\simeq0.0087$) satisfying PBH as dark matter and detectable GW signal. 
 We summarize the main findings of our analysis below:
\begin{itemize}
    \item Heavy neutrino-like dark fermions impact radiatively the hybrid inflation potential via quantum loop corrections to the inflaton field involved. Consequently, the predictions for CMB observables are affected due to the variation of the dark fermion mass m$_N$ (see \cref{parmsets}). The prediction for \textit{tensor-to-scalar} ratio $r\simeq0.0087$ and \textit{scalar spectral index} $n_s\simeq 0.964$, lies within current Planck bounds and testable in future CMB experiments like LiteBIRD, SO and CMB S4-Euclid.
    \item We estimate the power spectrum across all k-values which provides constraints on the dark fermion mass scale from the measurements of CMB spectral distortions (see \cref{fig:PS_k}). Along the valley, when the inflaton field becomes smaller than the critical value $\varphi_c$, the effective mass square of the waterfall field becomes negative. This gives rise to tachyonic instability and the power spectrum shoots up at small scales. The radiative corrections from the dark fermions along with the inflaton control the amplitude of the plateau and define the number of e-folds along the valley (see middle left of \cref{fig:PS_k}).
    \item Second-order tensor perturbations propagating as GWs that can be with amplitude $\Omega_{\rm GW}h^2$ $\sim 10^{-9}$ and peak frequency f $\sim$ 0.1 Hz by LISA and $\Omega_{\rm GW}h^2 \sim 10^{-11}$ and peak frequency of $\sim$ 10 Hz in ET in this model (see \cref{fig:Omegaf}). 
    \item Production of PBH of mass around $ 10^{-13} M_\odot$ as the sole DM candidate in the universe is proposed. This novel DM candidate is also a signature of the scale of dark fermion physics involving inflationary cosmology (see \cref{fig:f_pbh}). 
    \item We estimate fine-tuning in our model and found that it is around five orders of magnitude smaller than single field inflation and one order of magnitude smaller than other hybrid inflationary scenarios studied in the literature (see \cref{finetune}).
    \item For choice of benchmark point 3 (BP-3) in \cref{parmsets}, one finds that dark fermions can be a possible candidate for being a RHN which is responsible for the generation of SM neutrino mass via the seesaw mechanism and leaves imprints in the power spectrum and GW. Not only it can be a fractional DM of the universe but also be tested in future probes of the Nancy Grace Roman Space Telescope (see \cref{fig:f_pbh}). Reheating in such a case may proceed via a waterfall field decaying into RHN and is suitable for leptogenesis via subsequent decays of RHN. We find the reheating temperature $T_R\lesssim 5\times 10^{9}$ GeV that may explain matter-antimatter asymmetry leptogenesis, neutrino mass $m_N\simeq8.28\times 10^{11}$ GeV and the corresponding PBHs are of $10^{-9}\,M_\odot$.
\end{itemize} 
It has been well studied that the presence of non-Gaussianities affects the abundance of PBH formation quite a significant manner; see refs.~\cite{Bullock:1996at,Ivanov:1997ia,Yokoyama:1998pt,Hidalgo:2007vk,Byrnes:2012yx,Bugaev:2013vba,Young:2015cyn,Nakama:2016gzw,Ando:2017veq,Franciolini:2018vbk,Atal:2018neu,Passaglia:2018ixg,Atal:2019cdz,Atal:2019erb,Yoo:2019pma,Taoso:2021uvl,Kitajima:2021fpq,Escriva:2022pnz,Franciolini:2023agm,Choudhury:2023fwk,Choudhury:2023kdb} and along with PBH clustering, see refs.~\cite{Chisholm:2005vm,Young:2014ana,Young:2014oea,Tada:2015noa,Young:2015kda,Suyama:2019cst,Young:2019gfc}). It also affects the predictions of induced GW signals (see, e.g., refs.~\cite{Cai:2018dig,Unal:2018yaa,Yuan:2020iwf,Atal:2021jyo,Adshead:2021hnm,Garcia-Saenz:2022tzu,Abe:2022xur} and~\cite{Domenech:2021ztg} for a review on this topic). Since in our scenario the curvature perturbation is generated and enhanced during the waterfall transition, this non-Gaussianity may have an impact~\cite{Kawasaki:2015ppx} but it is beyond the scope of the present work and will be taken up in future publication; for generic discussions of $f_\text{NL}$ on SIGW and PBH see \cite{Li:2023xtl,Li:2023qua,Choudhury:2023kdb}. Recently, the quantum loop correction from the PBH-scale perturbation to the CMB-scale is being actively debated for the single-field inflationary scenario~\cite{Inomata:2022yte, Kristiano:2022maq, Riotto:2023hoz, Choudhury:2023vuj, Kristiano:2023scm, Riotto:2023gpm, Firouzjahi:2023aum, Motohashi:2023syh, Bhattacharya:2023ysp,Franciolini:2023lgy}. However, the possible No-Go theorem proposed in ref.~\cite{Kristiano:2022maq} would also be interesting to understand for the loop correction in hybrid inflation. We envisage that GW astronomy with the planned global network of GW detectors can make the dream of testing high-scale and fundamental BSM scenarios like seesaw scale and neutrino physics involving UV-completion and inflationary cosmology a reality.

\medskip

\section*{Acknowledgments}
We thank BCVSPIN school and the hospitality of Kathmandu and Tribhuvan Universities in Nepal where this project was initiated. We thank Qaisar Shafi for several discussions and useful clarifications during the completion of the project. We also thank Ioanna D. Stamou, Ahmad Moursy and Stefan Antusch for useful comments on the manuscript.


\appendix
\section{Appendix: Gravitational Wave Signal in Pulsar Timing Array}
\label{appGW}
\begin{figure}[H]
    \centering
    \includegraphics[width=0.471\linewidth,height=5.7cm]{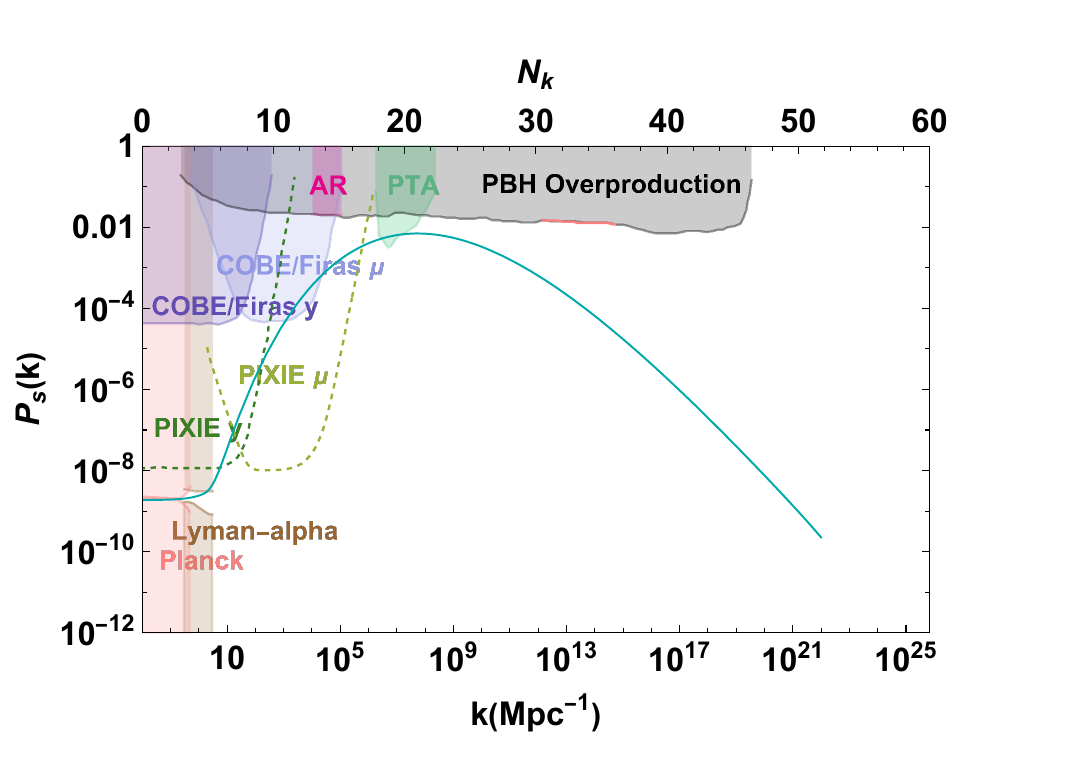}
    \quad
    \includegraphics[width=0.471\linewidth,height=5.7cm]{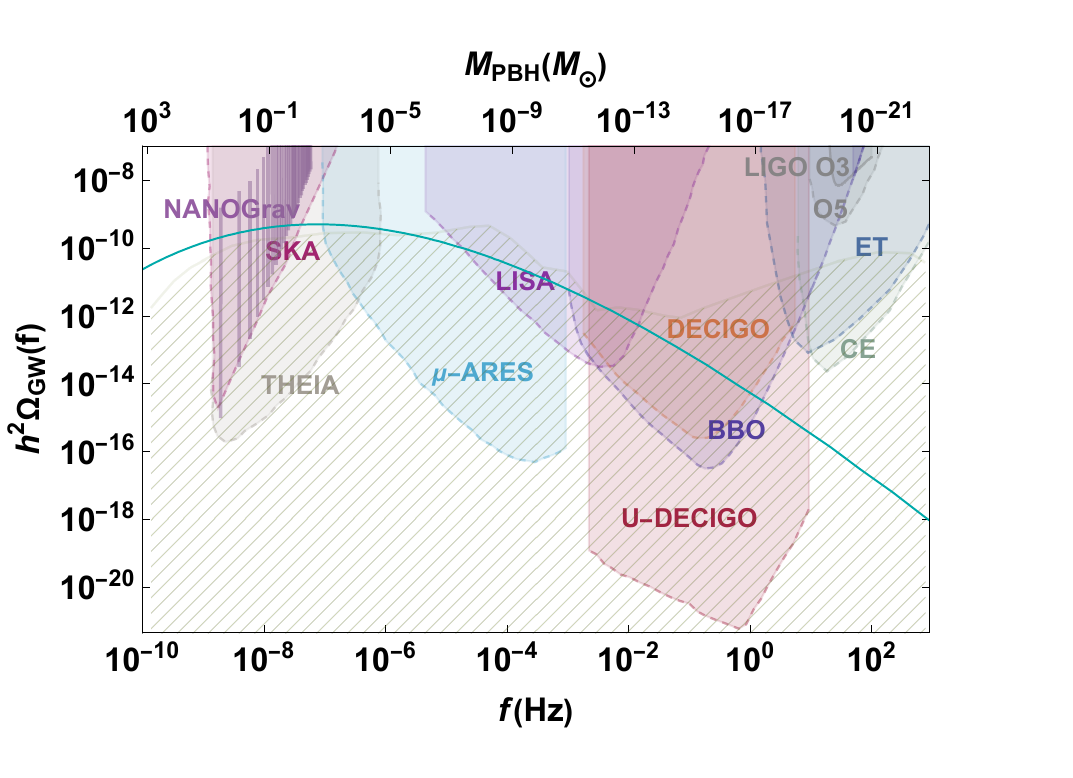}
    \caption{ \it  Left panel shows the power spectrum by solving the exact scalar perturbation equations for the BP given in \cref{parmsetsnano}. The shaded region corresponds to the constraints from the present (solid) and future (dashed) experiments. For further details on experimental bounds, see the main text in \cref{scalarpert}. The right panel indicates the energy density of GWs for \cref{Energdens} for the BPs given in \cref{parmsetsnano}.  The colored shaded regions indicate the sensitivity curves of present (solid boundaries) LIGO O3 \cite{KAGRA:2021kbb}, NANOGrav \cite{NANOGrav:2023gor} and future (dashed boundaries) LIGO O5, SKA \cite{Smits:2008cf}, THEIA \cite{Garcia-Bellido:2021zgu}, LISA \cite{Baker:2019nia}, $\mu$-ARES \cite{Sesana:2019vho}, BBO \cite{Corbin:2005ny}, U-DECIGO \cite{Yagi:2011wg, Kawamura:2020pcg}, CE \cite{Reitze:2019iox} and ET \cite{Punturo:2010zz} experiments. The hatched region shows the astrophysical background 
\cite{ghoshal2023traversing}.}
    \label{fig:PONANO}
\end{figure}

To test the model presented in this paper in the PTA band, the $n_s$ conflicts with the Planck 2018. Moreover, the power spectrum is also ruled out by COBE/Firas due to the steep enhancement at large scales, as shown in the left panel of \cref{fig:PONANO} for the BP in \cref{parmsetsnano}. This conflict can be encountered by including the effect of non-Gaussanity parameter $f_\text{NL}$, we leave this for future work\footnote{See ref.~\cite{Chen:2019xse} for SIGW satisfying NANOGrav measurements. For non-Gaussianity and secondary gravitational waves from PBH production in $\alpha$-attractor inflation, see~\cite{Rezazadeh:2021clf}.}.

\begin{table}[tbh!]
	  \fontsize{16pt}{15pt}
	\caption{ \it Benchmark points for model parameters in the PTA} 
	\centering 
	\begin{adjustbox}{max width=\columnwidth}
		\begin{tabular}{|M{2.0cm} |M{1.7cm} |M{2.5cm} |M{2.4cm} |M{2.6cm}|M{2.7cm}|M{1.9cm}|M{1.8cm}|M{1.7cm}|M{1cm}|M{1cm}|M{1.5cm}|}
			\hline
			\bf{Model}& \bf{$M/m_\text{Pl}$}&\bf{$m/m_\text{Pl}$}&\bf{$\lambda$}&\bf{$\kappa$}&\bf{$b$}&\bf{$\sqrt{\alpha}/m_\text{Pl}$}&\bf{$\varphi_i/m_\text{Pl}$}&\bf{$\psi_i/m_\text{Pl}$}
			\\ [0.6ex] 
			\hline\hline
			$\text{BP-PTA}$ &$1.47$  &$6.10\times 10^{-7}$  & $6.30\times 10^{-6}$ & $ 1.45\times 10^{-6} $ &$-3.0\times 10^{-6}$ &$0.83$ & $3.5$ & $0$\\
			\hline 
		\end{tabular}
	\end{adjustbox}
	\label{parmsetsnano}
\end{table}
\begin{table}[tbh!]
	 \fontsize{11pt}{15pt}
	\centering 
	\begin{adjustbox}{max width=\columnwidth}
		\begin{tabular}{|M{1.8cm}|M{1.8cm} |M{2.1cm} |M{1.8cm} |M{1.4cm}|M{1cm}|M{1cm}|M{1.5cm}|}
			\hline
			\bf{Model}& \bf{$\mu/m_\text{Pl}$}&\bf{$m_N/m_\text{Pl}$}&\bf{$y$}&\bf{$\phi_c/m_\text{Pl}$}&\bf{$N_k$}&\bf{$n_s$}&\bf{$r$}
			\\ [0.6ex] 
			\hline\hline
			$\text{BP-PTA}$ &$2.17 \times 10^{-6}$  &$1.15 \times 10^{-3}$  & $4.45\times 10^{-6}$ & $ 0.487 $ & $55$ & $1.097$ & $0.00527$ \\
			\hline 
		\end{tabular}
	\end{adjustbox}
\end{table}

\bibliography{Bibliography}
\bibliographystyle{JHEP}
\end{spacing}
\end{document}